\documentclass[journal]{IEEEtran}[10pt]
\IEEEoverridecommandlockouts
\usepackage{cite}
\usepackage{makecell}
\usepackage{amsmath,amssymb,amsfonts}
\usepackage{algorithmic}
\usepackage{bm}
\usepackage{graphicx}
\usepackage{subfigure}
\usepackage{makecell}
\newtheorem{theorem}{Theorem}
\newtheorem{lemma}{Lemma}

\usepackage{textcomp}
\usepackage{epstopdf}
\usepackage{multirow}
\usepackage{xcolor}

\def\BibTeX{{\rm B\kern-.05em{\sc i\kern-.025em b}\kern-.08em
		T\kern-.1667em\lower.7ex\hbox{E}\kern-.125emX}}
\usepackage{stfloats,lipsum} 

\begin{document}
\title{Enhancing Spatial Multiplexing and Interference Suppression for Near- and Far-Field Communications with Sparse MIMO}
	\author{
	\IEEEauthorblockN{Huizhi~Wang, Chao~Feng, Yong~Zeng,~\IEEEmembership{Senior Member, IEEE},
	Shi~Jin,~\IEEEmembership{Fellow, IEEE},	\\	Chau~Yuen,~\IEEEmembership{Fellow, IEEE},	
	Bruno~Clerckx,~\IEEEmembership{Fellow, IEEE}} and Rui~Zhang,~\IEEEmembership{Fellow, IEEE}
	\thanks{		
		Part of this work has been presented at the 2023 IEEE GLOBECOM Workshops, Kuala Lumpur, Malaysia, in Dec. 2023 \cite{b109}.

Huizhi Wang, Chao Feng, Yong Zeng and Shi Jin are with the National Mobile Communications Research Laboratory, Southeast University, Nanjing 210096, China. Yong Zeng is also with the Purple Mountain Laboratories, Nanjing 211111, China (E-mail: \{wanghuizhi, chao\_feng, yong\_zeng, jinshi\}@seu.edu.cn). (\emph{Corresponding author: Yong Zeng.})

Chau Yuen is with the School of Electrical and Electronic Engineering, Nanyang Technological University, 639798, Singapore (E-mail: chau.yuen@ntu.edu.sg).

Bruno Clerckx is with the Department of Electrical and Electronic Engineering, Imperial College London, London, SW7 2AZ, UK and with Silicon Austria Labs (SAL), Graz A-8010, Austria (E-mail: b.clerckx@imperial.ac.uk).

Rui Zhang is with School of Science and Engineering, Shenzhen Research Institute of Big Data, The Chinese University of Hong Kong, Shenzhen, Guangdong 518172, China (e-mail: rzhang@cuhk.edu.cn). He is also with the Department of Electrical and Computer Engineering, National University of Singapore, Singapore 117583 (E-mail: elezhang@nus.edu.sg).
}}

\maketitle
\begin{abstract}
	\textbf{Multiple-input multiple-output (MIMO) has been a key technology for wireless systems for decades. For typical MIMO communication systems, antenna array elements are usually separated by half of the carrier wavelength, thus termed as conventional MIMO. In this paper, we investigate the performance of multi-user MIMO communication, with sparse arrays at both the transmitter and receiver side, i.e., the array elements are separated by more than half wavelength. Given the same number of array elements, the performance of sparse MIMO is compared with conventional MIMO. On one hand, sparse MIMO has a larger aperture, which can achieve narrower main lobe beams that make it easier to resolve densely located users. Besides, increased array aperture also enlarges the near-field communication region, which can enhance the spatial multiplexing gain, thanks to the spherical wavefront property in the near-field region. On the other hand, element spacing larger than half wavelength leads to undesired grating lobes, which, if left unattended, may cause severe inter-user interference (IUI). To gain further insights, we first study the spatial multiplexing gain of the basic single-user sparse MIMO communication system, where a closed-form expression of the near-field effective degree of freedom (EDoF) is derived. The result shows that the EDoF increases with the array sparsity for sparse MIMO before reaching its upper bound, which equals to the minimum value between the transmit and receive antenna numbers. Furthermore, the scaling law for the achievable data rate with varying array sparsity is analyzed and an array sparsity-selection strategy is proposed. We then consider the more general multi-user sparse MIMO communication system. It is shown that sparse MIMO is less likely to experience severe IUI than conventional MIMO, especially when users are densely located, thanks to the non-uniform distribution of spatial angle difference among users. Finally, numerical results are provided to validate our theoretical analysis. }
\end{abstract}
\section{introduction}
Multiple-input multiple-output (MIMO) is a key transmission technology in contemporary wireless communication systems, thanks to its additional spatial diversity gain and multiplexing gain over single-antenna systems\cite{b19}. Over the past few decades, MIMO communications have been tremendously advanced from small MIMO in 4G to massive MIMO in 5G\cite{b21}. Looking forward towards 6G, there have been growing interests in the study of extremely large-scale MIMO (XL-MIMO) \cite{b2,b22,b23,b3,b24,b117}, where several hundreds or even thousands of antennas are considered. By deploying such extremely large number of antennas, XL-MIMO can realize large array aperture and thus significantly enhance its spatial resolution, which is beneficial to both communication and sensing\cite{b142}. Besides XL-MIMO, there are some other similar terminologies, including extremely large aperture array (ELAA) \cite{b132}, ultra-massive MIMO (UM-MIMO) \cite{b133}, and extremely large aperture massive MIMO (xMaMIMO)\cite{b29}. 

For typical MIMO communication systems, antenna elements are usually separated by half of the signal wavelength, which are termed {\emph {conventional MIMO}}. As the antenna size grows, conventional XL-MIMO faces practical challenges due to significantly increased deployment cost and computational complexity. An alternative method to enhance the total array aperture, yet without having to increase the number of array elements, is to use {\emph {sparse MIMO}}, whose array element spacing is larger than half wavelength\cite{b109,b149}. Loosely speaking, depending on whether adjacent array elements have equal separation or not, sparse MIMO can be classified into uniform and non-uniform sparse MIMO. One particular type of non-uniform sparse array is modular arrays, where the power scaling law and beam focusing patterns have been analyzed in\cite{b143}. Besides, non-uniform sparse arrays have been extensively investigated for localization and sensing systems\cite{b136,b129}. The key idea of sparse arrays for localization is to use $O(N)$ antennas to identify up to $O(N^2)$ targets  and further improve the accuracy of direction of arrival (DoA) estimation, by constructing virtual arrays, including \emph{difference co-array}\cite{b128} and \emph{sum co-array}\cite{b138,b139}. To this end, several non-uniform sparse MIMO structures have been proposed, e.g., nested array\cite{b126} and co-prime array\cite{b129}. However, the virtual co-arrays are usually created by calculating high-order cumulants of the received signal, which will cause distortion of the information symbols if such a technique is directly applied to the communication system. Besides, designing non-uniform sparse array is non-trivial since the tradeoff between mutual coupling and spatial resolution enhancement needs to be carefully considered. To gain the essential insights for communication, in this paper, we mainly focus on the basic uniform sparse MIMO with sparse arrays at both transmitter and receiver sides, whose value is greater than half signal wavelength.

In particular, we aim to answer the following fundamental question: Given the same number of array elements, whether uniform sparse MIMO can achieve better communication performance than the conventional MIMO? The answer to this question is not immediately clear for the following reasons. On one hand, compared with conventional counterpart, sparse MIMO will achieve narrower main lobe beams to attain finer spatial resolution, thanks to its larger total array aperture. Therefore, sparse MIMO is expected to better resolve densely located users, which is appealing for hot-spot communication scenario. On the other hand, sparse MIMO with inter-element spacing larger than half wavelength will suffer from undesired grating lobes, i.e., equally strong power will be radiated towards some other directions besides the desired main lobe direction. In this case, for communication systems, two users located within grating lobes of each other will suffer from severe inter-user interference (IUI).

For sparse MIMO, as the array sparsity further increases, the near-field region of the array will expand. As a result, users used to be located in the far-field region of conventional MIMO may enter the near-field region of sparse MIMO. Therefore, the performance of sparse MIMO considering the general near-field effect needs to be analyzed. Specifically, for conventional MIMO with a small number of array elements, the spatial multiplexing gain is limited in far-field region due to the uniform plane wave (UPW) propagation, and the degree-of-freedom (DoF) is only one for the line-of-sight (LoS) channel\cite{b137}. In contrast, for sparse MIMO with equal number of array elements at the transmitter and receiver, the spatial multiplexing gain may be significantly enhanced, thanks to the near-field non-uniform spherical wave (NUSW) property with non-linear phases and non-uniform amplitudes across the array elements\cite{b117}.

There are some preliminary works on the performance study of communication systems with sparse arrays. In the conference version of the current work\cite{b109}, we showed that communication systems with sparse array at the base station (BS) will be less susceptible to severe IUI compared to conventional array. This is because the higher-order grating lobes of sparse arrays do not necessarily cause strong IUI, thanks to the typically non-uniform distribution of the spatial angle difference between users. However, the work \cite{b109} only considered the single-input multiple-output (SIMO) setup and the analysis therein was only based on the conventional far-field UPW model. On the other hand, by considering spherical wave property, the possibility of realizing a higher channel matrix rank for LoS MIMO system has been studied in \cite{b146,b147,b148,b150,b151}, and the DoF performance for XL-MIMO systems has been investigated in \cite{b145,b108,b115}. To the best of our knowledge, the rigorous performance analysis of uniform sparse MIMO communication in terms of spatial multiplexing and interference suppression for both far-field and near-field communications has not been fully explored yet. The main contributions of this paper are summarized as follows:
\begin{itemize}
	\item First, we present the multi-user communication system with sparse MIMO configuration, where both the user equipment (UE) and BS are equipped with uniform sparse arrays with element-spacing larger than half signal wavelength. To gain useful insights, the single-user case is first considered, which can be viewed as a special case of the multi-user scenario. In this case, the performance of sparse MIMO communication is analyzed from the spatial multiplexing perspective. In order to obtain useful insights, we explore the relationship between the achievable data rate and the channel DoF for MIMO communication. In particular, we use the effective degree of freedom (EDoF)\cite{b108,b131} as a metric to assess the multiplexing performance.
	
	\item Next, we derive the closed-form expressions for EDoF in far-field and near-field scenarios, respectively, which are shown to be a piecewise function of array sparsity. The dependence of the closed-form EDoF on key system parameters, such as the number of array elements, the distance of the user from the center of BS, and the sparsity of the transmit/receive arrays is characterized. It is revealed that compared with the conventional MIMO, by using sparse MIMO with inter-element spacing greater than half signal wavelength, the EDoF can be increased greatly. Besides, our newly derived EDoF includes the conventional far-field result as a special case. Based on the fact that the EDoF for far-field communication in free-space LoS scenario is always one, a new criterion to distinguish near-field and far-field regions is derived, which is related to the number of array elements at the UE and BS, and the array sparsity. Furthermore, by exploring the scaling law of achievable data rate with different array sparsity, we propose an antenna selection strategy for sparse MIMO communication, in order to maximize the data rate. 
	
	\item Finally, for the general multi-user case, interference suppression for sparse MIMO is studied for far-field and near-field communications. It is revealed that for the far-field case, the interference mitigation performance of the single-cell multi-user system only depends on the sparsity of the array at the BS side, but is independent of the array sparsity at the user side. Furthermore, for the near-field case, our simulation results show that the users' sum rate increases significantly with the array sparsity at BS. It is also revealed that for far-field users, it is better to deploy sparse array at the BS, instead of at the UEs. For near-field users, the array sparsity at the UE and the BS needs to be carefully chosen, respectively. 
\end{itemize}

The rest of this paper is organized as follows. Section \uppercase\expandafter{\romannumeral2} introduces the general near-field model for multi-user sparse MIMO communication, with sparse arrays equipped at both the UE and BS. Section \uppercase\expandafter{\romannumeral3} studies the multiplexing enhancement for single-user sparse MIMO communication, where the scaling law of achievable data rate as well as the closed-form expression of EDoF is derived. Section \uppercase\expandafter{\romannumeral4} derives the sum rate of multi-user scenario for both far-field and near-field MIMO communications. Section \uppercase\expandafter{\romannumeral5} provides numerical results to validate our theoretical findings.  

\textit{Notations}: Lower and upper-case bold letters denote vectors and matrices, respectively. The $i$th element of a vector $\mathbf{z}$ is denoted by $z_i$. The $(m,n)$th element of a matrix $\mathbf{Z}$ is denoted by $\left[\mathbf{Z}\right]_{m,n}$. Besides, ${\bf{Z}}^H,$ $\det ({\bf{Z}}),$ ${\| {\bf{Z}} \|_F}$ and $\mathrm{tr} ({\bf{Z}})$ denote the conjugate transpose, determinant, Frobenius norm and trace of the matrix ${\bf{Z}}$, respectively. Finally, ${\| {\bf{z}} \|}$ refers to the Euclidean norm of the vector$\ {\bf{z}}\ $.
\section{system model}
\begin{figure}[htbp]
	\setlength{\abovecaptionskip}{-0.1cm}
	\setlength{\belowcaptionskip}{-0.3cm}
	\centerline{\includegraphics[width=0.5\textwidth]{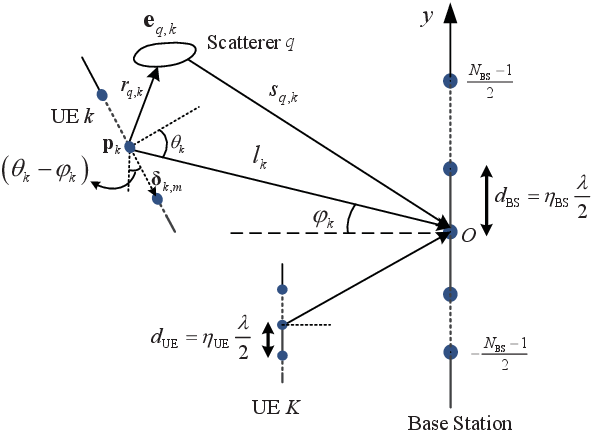}}
	\caption{Multi-user sparse MIMO communication, where the antenna separation at the BS and UE are $\eta_{\rm{BS}}\frac{\lambda}{2}$ and $\eta_{\rm{UE}}\frac{\lambda}{2}$, respectively.}
	\label{setup}
\end{figure}
As shown in Fig.\ref{setup}, we consider a multi-user sparse MIMO communication system with a BS and $K$ UEs, where uniform sparse arrays are equipped at both the BS and UEs. Specifically, the BS is equipped with a sparse uniform linear array (S-ULA) with ${N_{\rm{BS}}}$ elements, which are separated by distance $d_\mathrm{BS} = {\eta_\mathrm{BS}}\frac{\lambda}{2}$, where $\eta_{\rm{BS}}\geq 1$ and $\lambda$ is the signal wavelength. Besides, each user is equipped with ${N_{\rm{UE}}}$ antennas, with adjacent elements separated by $d_\mathrm{UE} = {\eta_\mathrm{UE}}\frac{\lambda}{2}$, where $\eta_{\rm{UE}}\geq 1$. Note that ${\eta_\mathrm{BS}}$ and ${\eta_\mathrm{UE}}$ account for the sparsity of arrays deployed at the BS and UEs, respectively. It is worth mentioning that our proposed system includes the conventional MIMO as a special case, by setting ${\eta_\mathrm{BS}}={\eta_\mathrm{UE}}=1$.
\begin{figure*}[hb] 
	\hrulefill
	\centering
	\begin{equation}
		\setlength\abovedisplayskip{2pt}
		\setlength\belowdisplayskip{2pt}
		\small
		\begin{aligned}
			l_{m,n}^k &= \left\| {{{\bf{p}}_{k,m}} - {{\bf{w}}_n}} \right\| = \sqrt {{{\left( {  {l_k}\cos \varphi_k + m{d_\mathrm{UE}}\sin (\theta_k -\varphi_k )} \right)}^2} + {{\left( {{l_k}\sin \varphi_k + m{d_\mathrm{UE}}\cos (\theta_k -\varphi_k ) - n{d_\mathrm{BS}}} \right)}^2}} \\
			&= {l_k}\sqrt {1 + \frac{\lambda }{{{l_k}}}\left( {m{\eta _{{\rm{UE}}}}\sin {\theta _k} - n{\eta _{{\rm{BS}}}}\sin {\varphi _k}} \right) + \frac{{{\lambda ^2}}}{{4l_k^2}}\left[ {{{(m{\eta _{{\rm{UE}}}})}^2} + {{(n{\eta _{{\rm{BS}}}})}^2} - 2mn{\eta _{{\rm{UE}}}}{\eta _{{\rm{BS}}}}\cos ({\varphi _k} - {\theta _k})} \right]} .
		\end{aligned}
		\label{distance}
	\end{equation}	
\end{figure*}
\begin{figure*}[htbp]
	\centering
	\subfigure[$\eta_{\rm{UE}}=4, \eta_{\rm{BS}}=2$.]{
		\includegraphics[width=5.7cm]{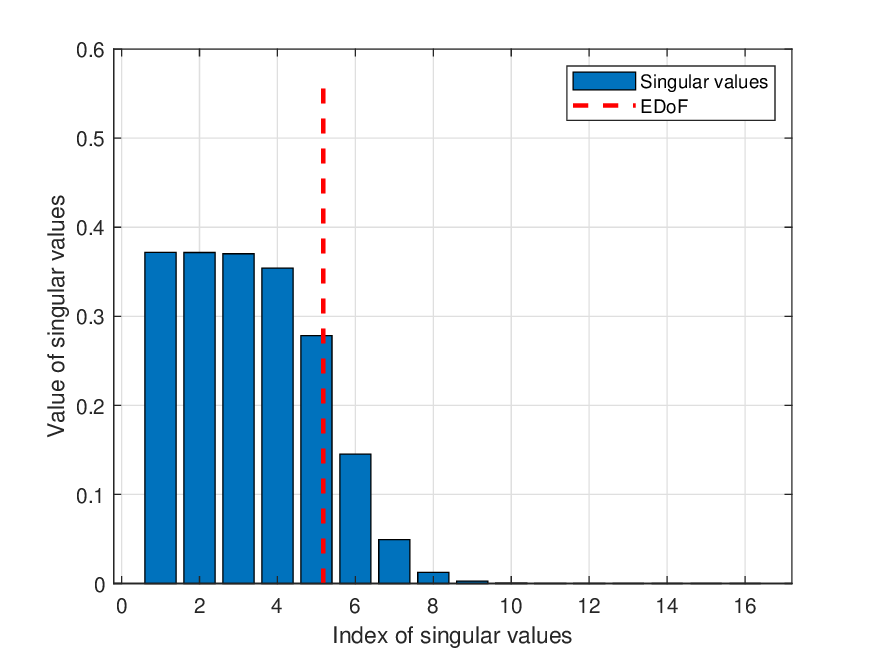}
	}
	\subfigure[$\eta_{\rm{UE}}=5, \eta_{\rm{BS}}=3$.]{
		\includegraphics[width=5.7cm]{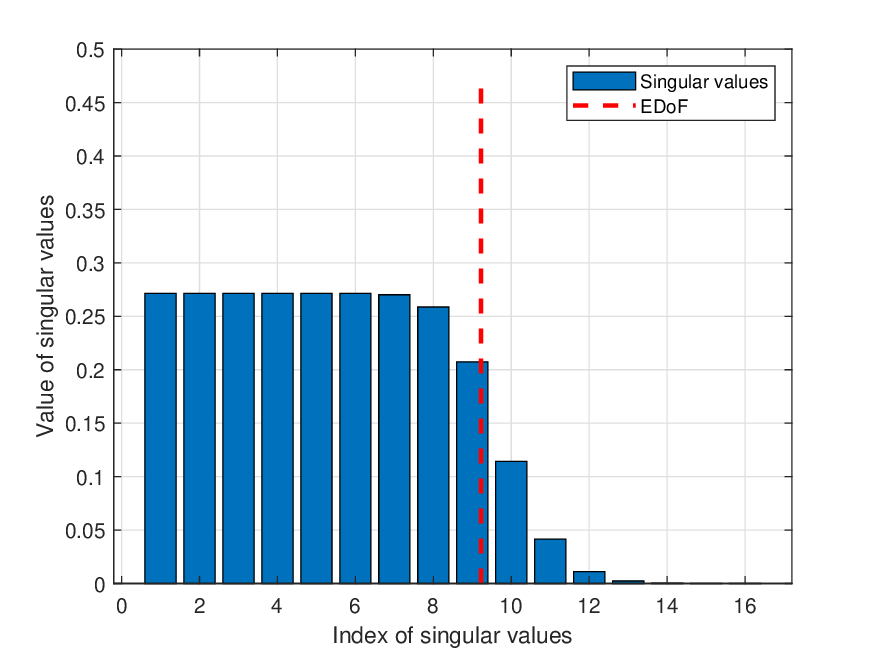}
	}
	\subfigure[$\eta_{\rm{UE}}=6, \eta_{\rm{BS}}=4$.]{
		\includegraphics[width=5.7cm]{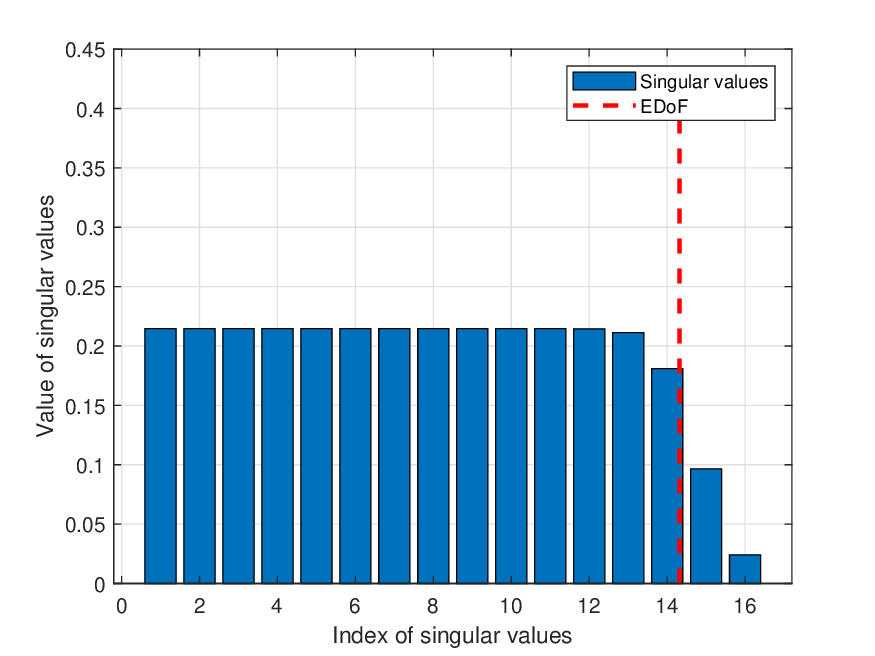}
	}
	\caption{Distribution of singular values for LoS sparse MIMO ($N_{\rm{UE}}=16, N_{\rm{BS}}=64$).}
	\label{distri}
\end{figure*}
~~For notational convenience, we assume that both ${N_{\rm{BS}}}$ and ${N_{\rm{UE}}}$ are odd numbers, and the total array aperture of the BS and UE can be expressed as $D_\mathrm{BS}= ({N_{\rm{BS}}}-1) d_\mathrm{BS}$ and $D_\mathrm{UE}= ({N_{\rm{UE}}}-1) d_\mathrm{UE}$, respectively. Furthermore, it is assumed that the S-ULA at the BS is placed along the $y$-axis and centered at the origin. Therefore, the location of the $n$th array element of the BS is given by ${\bf w}_n = [0,nd_\mathrm{BS}]^T,$ where $n\in\{0,\pm1,...,\pm({N_{\rm{BS}}}-1)/2\}$. The center of the S-ULA corresponding to the $k$th UE is denoted as ${\bf p}_k = [-l_k\cos\varphi_k,l_k\sin\varphi_k]^T,$ where $l_k$ is the distance between the array center of UE $k$ and the origin, and $\varphi_k\in [-\pi/2, \pi/2]$ denotes the direction of the central element at UE $k$ with respect to the BS's boresight. Therefore, for UE $k$, the location of its $m$th element can be expressed as ${\bf p}_{k,m} = {\bf p}_k+{\bm \delta}_{k,m}$ with $m\in\{0,\pm1,...,\pm({N_{\rm{UE}}}-1)/2\}$, and ${\bm \delta}_{k,m}$ denotes the location of the $m$th element relative to the reference point ${\bf p}_k$. Let $\theta_k\in [-\pi/2, \pi/2]$ denote the angle between $\mathbf{p}_k$ and the normal vector of the S-ULA at UE $k$, and ${\bm\delta}_{k,m}$ can be expressed as ${\bm \delta}_{k,m}=[{-m{d_\mathrm{UE}}\sin (\theta_k -\varphi_k )},{m{d_\mathrm{UE}}\cos (\theta_k -\varphi_k )}]^T$. The distance between the $m$th element of UE $k$ and the $n$th element of the BS is denoted as $l_{m,n}^k$, which can be further expressed as (\ref{distance}) shown at the bottom of this page. Note that (\ref{distance}) is the exact distance expression that can be degenerated to the conventional far-field UPW model by using first-order Taylor approximation of $l_{m,n}^k$. With the uniform spherical wave (USW) model, the channel coefficients of the LoS channel matrix component ${\bf H}_k^{\mathrm{LoS}}\in \mathbb{C}^{{N_{\rm{BS}}}\times {N_{\rm{UE}}}}$ between UE $k$ and the BS are given by\cite{b117}
\begin{equation}
	\setlength\abovedisplayskip{2pt}
	\setlength\belowdisplayskip{2pt}
	\begin{aligned}
	{[{\bf H}_k^{\mathrm{LoS}}]_{m,n}} = {\beta _k}{e^{ - j\frac{{2\pi }}{\lambda }(l_{m,n}^k - {l_k})}},
	\end{aligned}
	\label{los}
\end{equation}
where ${\beta _k}$ denotes the reference complex-valued channel gain. We can easily obtain $ \left\| {\bf{H}}_k^{\mathrm{LoS}} \right\|_F^2 = {{N_{\rm{BS}}}{N_{\rm{UE}}}|\beta_k|^2 }.$

Denote by $Q_k$ the number of multi-paths of UE $k$, and ${\bf e}_{q,k}$ the scatterer's location of path $q$ for UE $k$. Thus, the NLoS channel component of UE $k$ can be expressed as 
\begin{equation}
	\setlength\abovedisplayskip{2pt}
	\setlength\belowdisplayskip{2pt}
	\begin{aligned}
{\bf{H}}_k^{{\rm{NLoS}}} = \sqrt {\frac{{{\rho _k}}}{{{Q_k}}}} \sum\limits_{q = 1}^{{Q_k}} {{\alpha _{q,k}}{e^{ - j\frac{{2\pi }}{\lambda }r_{q,k} + j{\psi _q}}}{\bf{b}}({{\bf{e}}_{q,k}}){\bf{a}}^H({{\bf{e}}_{q,k}})},
	\end{aligned}
	\label{nlos_channel}
\end{equation}
where ${\alpha _{q,k}}$ denotes the complex channel coefficient related to the $q$th scatterer for UE $k$ with ${Q_k^{ - 1}}\sum\nolimits_{q = 1}^{Q_k} {\mathbb{E}[\alpha_{q,k}^2]}  = 1$. $ {\rho _k} \triangleq \sum\limits_{q = 1}^{{Q_k}} {\frac{{{\lambda ^2}{{\tilde \rho }_q}}}{{{{(4\pi )}^3}r_{q,k}^2s_{q,k}^2}}} $ denotes the total power of the NLoS channel paths at the reference point, where the parameter ${{{\tilde \rho }_q}}$ incorporates the radar cross section (RCS) of scatterer $q$. $r_{q,k}$ and $s_{q,k}$ denote the distance between the $q$th scatterer and the center of UE $k$ and the origin, respectively. $\psi_q$ denotes the additional
phase shift brought by scatterer $q$. 

For the uplink communication, ${{{\bf{a}}}_k({{\bf{e}}_{q,k}})}$ and ${{{\bf{b}}}({{\bf{e}}_{q,k}})}$ denote the near-field transmit and receive response vectors with respect to scatterer $q$, respectively, whose entries are given by
\begin{equation}
	\setlength\abovedisplayskip{2pt}
	\setlength\belowdisplayskip{2pt}
	\begin{aligned}
	&{\left[ {{{\bf{a}}}_k({{\bf{e}}_{q,k}})} \right]_m} =  {{e^{ - j\frac{{2\pi }}{\lambda }\left( {r_{q,k,m} - r_{q,k}} \right)}}}, \\
	&{\left[ {{{\bf{b}}}({{\bf{e}}_{q,k}})} \right]_n} =  {{e^{ - j\frac{{2\pi }}{\lambda }\left( {{s_{q,k,n}} - {s_{q,k}}} \right)}}},
	\end{aligned}
	\label{steer vector}
\end{equation}
where $r_{q,k,m} \triangleq \left\| {{{\bf{e}}_{q,k}} - {{\bf{p}}_{k,m}}} \right\|$ and $s_{q,k,n} \triangleq \left\| {{{\bf{e}}_{q,k}} - {{\bf{w}}_n}} \right\|$ account for the distance between the $q$th scatterer and the $m$th and $n$th elements of the $k$th UE and BS, respectively.
Therefore, the multi-path near-field channel matrix for UE $k$ can be expressed as\cite{b117}
\begin{equation}
	\setlength\abovedisplayskip{2pt}
	\setlength\belowdisplayskip{2pt}
	\begin{aligned}
{{\bf{H}}_k} = \sqrt {\frac{F_k}{{F_k + 1}}} {\bf{H}}_k^{{\rm{LoS}}} + \sqrt {\frac{1}{{F_k + 1}}} {\bf{H}}_k^{{\rm{NLoS}}},
	\end{aligned}
	\label{channel}
\end{equation}
where $F_k$ is the factor to present the relative power between the LoS component and the NLoS multipath components for UE $k$. 
The received signal at the BS can be obtained as
\begin{equation}
	\setlength\abovedisplayskip{1pt}
	\setlength\belowdisplayskip{2pt}
	\begin{aligned}
		{{\bf{y}}} = \sum\limits_{k = 1}^K {{{\bf{H}}_k}} {\bf x}_k + {\bf{z}},
	\end{aligned}
	\label{receive}
\end{equation}
where ${\bf x}_k$ is the transmit signal from UE $k$, and$\ {\bf z} \sim \mathcal{CN}(\mathbf 0,{\sigma ^2\mathbf{I}_{N_{\mathrm{BS}}}})\ $denotes the additive white Gaussian noise (AWGN).

\section{Single-user Sparse MIMO}\label{single}
To obtain useful insights about the scaling law of achievable data rate in terms of the array sparsity, we first consider the single user case with $K=1$, so that the subscript $k$ is omitted in this section. 
Let the transmitted signal by the UE be ${\bf x} \sim \mathcal{CN}(\mathbf 0,{\bf Q})$, where ${\bf Q}=E[{{\bf{x}}}{\bf{x}}^H]\in \mathbb{C}^{N_{\rm{UE}} \times N_{\rm{UE}}}$ is the transmit covariance matrix, with ${\rm{tr}}\left( {\bf{Q}} \right)\le P$, and $P$ is the maximum allowable power of the UE. The achievable data rate of a single-user MIMO communication system is~\cite{b111}
\begin{equation}
	\setlength\abovedisplayskip{2pt}
	\setlength\belowdisplayskip{2pt}
	\begin{aligned}
R = {\log _2}\det \left( {{{\bf{I}}_{N_{\mathrm{BS}}}} + \frac{1}{{{\sigma ^2}}}{\bf{HQ}}{{\bf{H}}^H}} \right).
	\end{aligned}
	\label{def}
\end{equation}
Note that $R$ depends on the transmit covariance matrix $\bf Q$, which is designed depending on whether the channel state information at the transmitter (CSIT) is available or not \cite{b111}. Therefore, the property of $R$ will be analyzed by considering different $\mathbf{Q}$ in the following.

{\bf {Without CSIT:}} The lack of CSIT tends to hinder the optimization of transmit power allocation and beamforming design at the UE. In this case, one may set ${\bf{Q}} = \frac{P}{{N_{\rm{UE}}}}{{\bf{I}}_{N_{\rm{UE}}}},$ which means the total power $P$ is equally allocated to each transmit antenna. Based on this, the achievable data rate in (\ref{def}) becomes 
\begin{equation}
	\setlength\abovedisplayskip{2pt}
	\setlength\belowdisplayskip{2pt}
	\begin{aligned}
	R = {\log _2}\det \left( {{{\bf{I}}_{N_{\mathrm{BS}}}} + \frac{P}{{{N_{\rm{UE}}}{\sigma ^2}}}{\bf{H}}{{\bf{H}}^H}} \right).
	\end{aligned}
	\label{without csit}
\end{equation}

{\bf {With CSIT:}} If the channel matrix $\bf{H}$ is known at the UE, we can first obtain independent single-input single-output (SISO) channels from such a MIMO system by performing the singular value decomposition (SVD). Specifically, the channel matrix $\bf H$ can be further written as ${\bf H = U\Sigma V}^H$, where the $N_{\rm{BS}}\times N_{\rm{BS}}$ matrix $\bf U$ and the $N_{\rm{UE}}\times N_{\rm{UE}}$ matrix $\bf V$ are unitary matrices and $\bf \Sigma$ is an $N_{\rm{BS}}\times N_{\rm{UE}}$ matrix containing singular values $\sigma_i$'s of $\bf H$ along its diagonal and zeros otherwise. Let $r_H$ denote the number of positive $\sigma_i$'s, which also refers to the rank of $\bf H$. Therefore, the MIMO channel can be decomposed into ${r_H}$ parallel SISO channels with the channel gains $\sigma_i, i=1,...,{r_H}$. The achievable data rate in (\ref{def}) can be further expressed as
\begin{equation}
	\setlength\abovedisplayskip{2pt}
	\setlength\belowdisplayskip{2pt}
	\begin{aligned}
	R = \sum\limits_{i = 1}^{r_H}  {{{\log }_2}\left( {1 + \frac{P_i}{\sigma^2}\sigma _i^2} \right)}, 
	\end{aligned}
	\label{single_rate_1}
\end{equation}
where $P_i$ denotes the power allocated to the $i$th channel, which satisfies $\sum\limits_{i = 1}^{{r_H}} {{P _i}}  \le P$. Furthermore, (\ref{single_rate_1}) can be maximized by applying the optimal water-filling power allocation\cite{b112}. Thus, the resulting data rate is given by\cite{b111}
\begin{equation}
	\setlength\abovedisplayskip{2pt}
	\setlength\belowdisplayskip{2pt}
	\begin{aligned}
R = \sum\limits_{{\gamma _i} \ge {\gamma _0}} {{{\log }_2}\left( {\frac{{{\gamma _i}}}{{{\gamma _0}}}} \right)},
	\end{aligned}
	\label{water-filling}
\end{equation}
where $\gamma_i\triangleq\sigma_i^2P/\sigma^2$ is the SNR associated with the $i$th channel at full power, and $\gamma_0$ is the cutoff value based on the power allocation constraint. On the other hand, when equal power allocation is applied to all the subchannels, i.e., $P_i=\frac{P}{r_H},$ the achievable data rate in (\ref{single_rate_1}) becomes \cite{b113}
\begin{equation}
	\setlength\abovedisplayskip{2pt}
	\setlength\belowdisplayskip{2pt}
	\begin{aligned}
R = \sum\limits_{i = 1}^{r_H} {{{\log }_2}\left( {1 + \frac{P}{{{r_H}{\sigma ^2}}}\sigma _i^2} \right)}.
	\end{aligned}
	\label{csit_uni}
\end{equation}
Furthermore, when the MIMO system is ideal and all the subchannels share similar characteristics, i.e., $\sigma_1\approx...\approx\sigma_{r_H}$, (\ref{csit_uni}) reduces to\cite{b114}
\begin{equation}
	\setlength\abovedisplayskip{2pt}
	\setlength\belowdisplayskip{2pt}
	\begin{aligned}
R = r_H {\log _2}\left( {1 + \frac{P }{r_H \sigma^2}\sigma _1^2} \right).
	\end{aligned}
	\label{single_rate_2}
\end{equation}

In practice, (\ref{single_rate_2}) is unreachable, because $\sigma_1\approx...\approx\sigma_{r_H}$ does not always hold and it is challenging to obtain the closed-form expression for ${r_H}$. Fortunately, an alternative metric, termed EDoF, can be used to approximate $r_H$. Specifically, the EDoF represents the equivalent number of sub-channels, which is given by \cite{b108}
\begin{equation}
	\setlength\abovedisplayskip{2pt}
	\setlength\belowdisplayskip{2pt}
	\begin{aligned}
\varepsilon = {\left( {\frac{{{\rm{tr}}({\bf{H}}{{\bf{H}}^H})}}{{{{\left\| {{\bf{H}}{{\bf{H}}^H}} \right\|}_F}}}} \right)^2}.
	\end{aligned}
	\label{edof_Def}
\end{equation}
As shown in Fig.~\ref{distri}, $r_H$ can be viewed as a critical threshold within which the value of singular values of $\bf H$ remains identical approximately. Besides, the exact value of $\sigma_1^2$ in (\ref{single_rate_2}) is still unknown, which can be approximated by the average of all the singular values as $\sigma _1^2 \approx \frac{{\left\| {\bf{H}} \right\|_F^2}}{\varepsilon }$\cite{b131}. 
To gain more insights, we first consider the LoS-dominant channel, i.e., $F\gg 1$ in (\ref{channel}). Therefore, by replacing $r_H$ with $\varepsilon$, a more practical expression of data rate can be written as
 \begin{equation}
 	\setlength\abovedisplayskip{2pt}
 	\setlength\belowdisplayskip{2pt}
 	\begin{aligned}
R \approx \varepsilon {\log _2}\left( {1 + \frac{{P\left\| {\bf{H}} \right\|_F^2}}{{{\sigma ^2}{\varepsilon ^2}}}} \right) = \varepsilon {\log _2}\left( {1 + \frac{{{N_{\rm{BS}}}{N_{\rm{UE}}}\bar P}}{{{\varepsilon ^2}}}} \right),
 	\end{aligned}
 	\label{single_rate_3}
 \end{equation}
where $\bar P \triangleq |\beta {|^2}\frac{P}{{{\sigma ^2}}}$ denotes the receive SNR before beamforming. Note that in (\ref{single_rate_3}), the identity $ \left\| {\bf{H}} \right\|_F^2 = {{N_{\rm{BS}}}{N_{\rm{UE}}}|\beta|^2 }$ is used. Let $C ={{{N_{\rm{BS}}}{N_{\rm{UE}}}\bar{P} }}$ and $x=\frac{1}{\varepsilon}$. Then, (\ref{single_rate_3}) can be rewritten as 
\begin{equation}
	\setlength\abovedisplayskip{2pt}
	\setlength\belowdisplayskip{2pt}
	\begin{aligned}
	R(x) = \frac{{1 }}{x}{\log _2}\left( {1 + C{x^2}} \right).
	\end{aligned}
	\label{single_rate_x}
\end{equation}
For convenience of analyzing $R(x)$, we define $\underline{N} \triangleq \min({N_{\rm{UE}}},{N_{\rm{BS}}})$, $\overline{N} \triangleq \max({N_{\rm{UE}}},{N_{\rm{BS}}})$, $\underline{\eta} \triangleq \min({\eta_{\rm{UE}}},{\eta_{\rm{BS}}})$ and $\overline{\eta} \triangleq \max({\eta_{\rm{UE}}},{\eta_{\rm{BS}}})$ in the following.
\subsection{Analysis of $R$}
\begin{theorem}\label{datarate}
	For single-user sparse MIMO communications, the achievable data rate $R(x)$ first increases and then decreases with $x>0$, and achieves its maximum value when ${x_{{\rm{opt}}}} = \sqrt {\frac{{{e^{W\left( - \frac{2}{{{e^2}}}\right) + 2}} - 1}}{C}}  \approx \frac{2}{{\sqrt C }}$, where $W(\cdot)$ is the Lambert $W$ function, satisfying $z = W(z){e^{W(z)}}$, $\forall z$.
\begin{IEEEproof}
	Please refer to Appendix A.
\end{IEEEproof}
\end{theorem}
Note that since $\varepsilon\in [1,\underline{N}]$, we can obtain $x\in [\frac{1}{\underline{N}},1]$. Therefore, ${x_{{\rm{opt}}}}$ may not be always attained, and thus three cases are considered in the following.

\begin{lemma}\label{lowsnr}
For $x_{\rm{opt}}>1$, the achievable data rate will increase with $x$, and ${R_{\max }}$ can be expressed as
\begin{equation}
	\begin{aligned}
		{R_{\max }} = R(1) = {\log _2}(1 + C).
	\end{aligned}
	\label{high}
\end{equation}
\end{lemma}
Since ${x_{{\rm{opt}}}} \approx \frac{2}{{\sqrt C }}$ and $C ={\underline{N}}{\overline{N}}\bar{P} ,$ $x_{\rm{opt}}>1$ in Lemma \ref{lowsnr} can be written as $\bar P <  \frac{4}{{\underline{N}\overline{N}}}$. In this case, the conventional compact array with $\varepsilon=1$ is the optimal. This is reasonable since there is no need to take advantage of the multiplexing gain when the SNR is low.

\begin{lemma}\label{highsnr}
For $x_{\rm{opt}}<\frac{1}{\underline{N}}$, the achievable data rate tends to decrease as $x$ becomes large. In this case, the maximum value of $R(x)$ is
	\begin{equation}
		\begin{aligned}
			{R_{\max }} = R\left( {\frac{1}{{\underline{N}}}} \right) = \underline{N}{\log _2}\left( {1 + \frac{C}{{{{\underline{N}}^2}}}} \right).
		\end{aligned}
		\label{low}
	\end{equation}
\end{lemma}
Lemma \ref{highsnr} indicates that for $\bar P >  \frac{4\underline{N}}{{\overline{N}}}$, the achievable date rate can be improved by increasing the EDoF. Thus, the maximum value of the EDoF $\varepsilon=\underline{N}$ maximizes the achievable data rate.

\begin{lemma}\label{medsnr}
For ${x_{{\rm{opt}}}}\in[\frac{1}{\underline{N}},1]$, i.e., $\bar P \in \left[ {\frac{4}{{\underline{N}\overline{N}}},\frac{4\underline{N}}{{\overline{N}}}} \right],$ the achievable data rate of single-user MIMO system can be maximized when $\varepsilon  \approx \frac{{\sqrt {\underline{N}\overline{N}\bar P} }}{2}.$ 
\end{lemma}

Based on Lemmas \ref{lowsnr} to \ref{medsnr}, the maximum achievable data rate can be written as
\begin{equation}
	\setlength\abovedisplayskip{2pt}
	\setlength\belowdisplayskip{2pt}
	\begin{aligned}
		{R_{\max }} = \begin{cases}
			{\log _2}(1 + \underline{N}\overline{N}\bar P),&\bar P < \frac{4}{\underline{N}\overline{N}}\\
			\underline{N}{\log _2}\left( {1 + \frac{{\overline N\bar P}}{\underline{N}}} \right),&\bar P > \frac{4\underline{N}}{{\overline N}}\\
			\frac{{\sqrt {\underline{N}\overline{N}\bar P} }}{2}{\log _2}5,&\bar P \in \left[ \frac{4}{{\underline{N}\overline{N}}},\frac{4\underline{N}}{{\overline N}} \right].
		\end{cases}
	\end{aligned}
\label{snr_all}
\end{equation}
Note that similar results in (\ref{snr_all}) can be seen in \cite{b150,b151}. However, while \cite{b150,b151} mainly focus on the optimal antenna rotation to achieve the maximum data rate, we will reveal how the sparsity of the antenna arrays, given a fixed antenna orientation, can increase the communication rate. Lemma \ref{lowsnr} to Lemma \ref{medsnr} show that for single-user near-field MIMO communications, the achievable data rate can be maximized with different values for $\varepsilon$ under different receive SNRs and antenna configurations. However, it is still unclear how the achievable data rate varies with the array sparsity $\underline{\eta}$ and $\overline{\eta}$. Therefore, in the following, we attempt to derive the closed-form expressions of $\varepsilon$ in terms of the array sparsity and gain useful insights into the relationship between them. 
\subsection{Closed-form expressions of $\varepsilon$}
Based on the definition of the EDoF in (\ref{edof_Def}), the generic expression of the near-field EDoF can be obtained as 
\begin{equation}
	\setlength\abovedisplayskip{2pt}
	\setlength\belowdisplayskip{2pt}
	\begin{aligned}
		{\varepsilon} = \frac{{\left\| {\bf{H}} \right\|_F^4}}{{{\rm{tr}}({{\bf{R}}^2})}} = \frac{{\underline{N}^2}{\overline{N}^2}|\beta|^4}{{\sum\limits_{n = 1}^{\underline{N}} {\sum\limits_{n' = 1}^{\underline{N}} {{{\left| {{R_{n',n}}} \right|}^2}} } }},
	\end{aligned}
	\label{epison_2}
\end{equation}
where ${\bf{R}}$ is the channel correlation matrix, given by\cite{b147}
\begin{equation}
	\setlength\abovedisplayskip{2pt}
	\setlength\belowdisplayskip{1pt}
	\begin{aligned}
		\bf R = \begin{cases}
			{\bf{H}}{{\bf{H}}^H},& {N_{\rm{UE}}}>{N_{\rm{BS}}} \\
			{{\bf{H}}^H}{{\bf{H}}},& {N_{\rm{UE}}}\le{N_{\rm{BS}}},
		\end{cases}
	\end{aligned}
\label{correlation}
\end{equation}
and
\begin{equation}
	\setlength\abovedisplayskip{2pt}
	\setlength\belowdisplayskip{2pt}
	\begin{aligned}
		{\left| {{R_{n',n}}} \right|^2} = \begin{cases} 
			{\left| {|\beta|^2 \sum\limits_{i = 1}^{N_{\rm{UE}}} {{e^{ - j\frac{{2\pi }}{\lambda }({l_{n',i}} - {l_{n,i}})}}} } \right|^2}, &{N_{\rm{UE}}}>{N_{\rm{BS}}} \\
			{\left| {|\beta|^2 \sum\limits_{i = 1}^{N_{\rm{BS}}} {{e^{ - j\frac{{2\pi }}{\lambda }({l_{i,n'}} - {l_{i,n}})}}} } \right|^2}, &{N_{\rm{UE}}}\le{N_{\rm{BS}}}.
			\end{cases}
	\end{aligned}
	\label{R}
\end{equation}
In order to further obtain the closed-form expression of ${\left| {{R_{n',n}}} \right|^2}$, far-field and near-field scenarios are separately analyzed. Note that the classic MIMO Rayleigh distance is given by\cite{b122}:
\begin{equation}
	\setlength\abovedisplayskip{2pt}
	\setlength\belowdisplayskip{2pt}
	\begin{aligned}
{l_{{\rm{Ray}}}} = 2\frac{{{{\left( {{D_{{\rm{UE}}}} + {D_{{\rm{BS}}}}} \right)}^2}}}{\lambda } \approx \frac{\lambda }{2}{\left( {{\overline N\overline \eta } + \underline{N}\underline{\eta}} \right)^2}.
	\end{aligned}
	\label{rayleigh}
\end{equation}
\subsubsection{Far-field scenario ($l > {l_{{\rm{Ray}}}}$)}\label{far-field}
\begin{theorem}\label{far}
For far-field free-space LoS MIMO communications, we obtain $\varepsilon=1$ for arbitrary antenna configuration.
\begin{IEEEproof}
	The distance $l_{m,n}$ in (\ref{distance}) reduces to 
	\begin{equation}
		\begin{aligned}
			{l_{m,n}} \approx R + m{d_\mathrm{UE}}\sin \theta  - n{d_\mathrm{BS}}\sin \varphi,
		\end{aligned}
	\label{r_far}
	\end{equation}
where the first-order Taylor approximation is used. By substituting (\ref{r_far}) into (\ref{R}), we have ${l_{n',i}} - {l_{n,i}} = \left( {n' - n} \right){d_{{\rm{UE}}}}\sin \theta ,$ and ${l_{i,n'}} - {l_{i,n}} = \left( {n-n'} \right){d_{{\rm{BS}}}}\sin \varphi,$ which are both independent of the summation index $i$. Therefore, ${\left| {{R_{n',n}}} \right|^2}$ in (\ref{R}) reduces to ${\left| {{R_{n',n}}} \right|^2}=\left| \beta  \right|^4{\overline{N}}^2$, and $\varepsilon$ can be expressed as $\varepsilon  = \frac{{{{\underline{N}^2}}{{\overline{N}^2}}|\beta {|^4}}}{{\sum\limits_{n = 1}^{\underline{N}} {\sum\limits_{n' = 1}^{\underline{N}} {{{\overline{N}^2}}|\beta {|^4}} } }} = 1.$
\end{IEEEproof}
\end{theorem}
Theorem \ref{far} shows that for the far-field scenario, the EDoF will not be affected by the sparsity of antenna configuration.

\subsubsection{Near-field scenario ($l \le {l_{{\rm{Ray}}}}$)}\label{near-single}
In this case, the more general scenario is considered where the UE is located at the near-field region of the BS, and the expression of $l_{m,n}$ in (\ref{distance}) can be expressed as
	\begin{equation}
	\setlength\abovedisplayskip{2pt}
	\setlength\belowdisplayskip{2pt}
	\begin{aligned}
	{l_{m,n}} &\mathop  \approx \limits^{(a)} l + \frac{\lambda }{2}\left( {m{\eta _{{\rm{UE}}}}\sin \theta  - n{\eta _{{\rm{BS}}}}\sin \varphi } \right) + \frac{{{{(\lambda m{\eta _{{\rm{UE}}}}\cos \theta )}^2}}}{{8l}}\\
&	 + \frac{{{\lambda ^2}}}{{8l}}{(n{\eta _{{\rm{BS}}}}\cos \varphi )^2} - \frac{{{\lambda ^2}}}{{4l}}mn{\eta _{{\rm{UE}}}}{\eta _{{\rm{BS}}}}\cos \varphi \cos \theta ,
	\end{aligned}
	\label{near-distance}
\end{equation}	
where $(a)$ follows by the second-order Taylor approximation $\sqrt{1+x}\approx1+\frac{1}{2}x-\frac{1}{8}x^2$, and the corresponding ${\left| {{R_{n',n}}} \right|^2}$ becomes
\begin{equation}
	\setlength\abovedisplayskip{2pt}
	\setlength\belowdisplayskip{2pt}
	\small
	\begin{aligned}
		{\left| {{R_{n',n}}} \right|^2}  \approx  |\beta {|^4}\frac{{{{\sin }^2}\left( {\frac{{\lambda \pi (n - n')\eta \cos \nu }}{{4l}}{\overline{N}}} \right)}}{{{{\sin }^2}\left( {\frac{{\lambda \pi (n - n')\eta \cos \nu }}{{4l}}} \right)}},
	\end{aligned}
	\label{near-R}
\end{equation}
where $\cos \nu  \triangleq \cos \theta \cos \varphi$ and $\eta\triangleq{\eta_\mathrm{BS}}{\eta_\mathrm{UE}}$.
Therefore, $\varepsilon$ can be expressed as
\begin{equation}
	\setlength\abovedisplayskip{2pt}
	\setlength\belowdisplayskip{2pt}
	\begin{aligned}
\varepsilon  = \frac{{{\underline{N}^2}}{{\overline{N}^2}}}{{\sum\limits_{n = 1}^{\underline{N}} {\sum\limits_{n' = 1}^{\underline{N}} {\frac{{{{\sin }^2}\left( {\frac{{\lambda \pi (n - n')\eta \cos \nu }}{{4l}}{\overline{N}}} \right)}}{{{{\sin }^2}\left( {\frac{{\lambda \pi (n - n')\eta \cos \nu }}{{4l}}} \right)}}} } }} = \frac{{{{\underline{N}^2}}{{\overline{N}^2}}}}{{\sum\limits_{n = 1}^{\underline{N}} {\sum\limits_{n' = 1}^{\underline{N}} {{f_\eta }(n - n')} } }},
	\end{aligned}
	\label{closed-form}
\end{equation}
where ${f_\eta }(\Delta ) \triangleq \frac{{{{\sin }^2}\left( {\frac{{\lambda \pi \eta {\overline{N}}\cos \nu }}{{4l}}\Delta } \right)}}{{{{\sin }^2}\left( {\frac{{\lambda \pi \eta \cos \nu }}{{4l}}\Delta } \right)}}$ with $ \Delta\in {{\rm{S}}_{{\rm{diff}}}} = \left\{ {{n_i} - {n_j},1 \le n_i,n_j \le {\underline{N}}} \right\}.$ Note that ${f_\eta }(\Delta )$ owns a similar form with the channel correlation of two UEs with spatial separation $\Delta$. Following similar analysis as \cite{b109}, we can obtain several properties of the function ${f_\eta }(\Delta )$ as follows:

\emph{Beamwidth of main lobe:} The null points of the main lobe for $f_\eta(\Delta)$ can be obtained by letting $\frac{{\lambda \pi \eta {\overline{N}}}\cos \nu}{{4l}}\Delta  =  \pm \pi $. Therefore, the null-to-null beamwidth of the main lobe is
\begin{equation}
	\setlength\abovedisplayskip{1pt}
	\setlength\belowdisplayskip{1pt}
	\begin{aligned}
		BW =  \frac{{8l}}{{\lambda \eta {\overline{N}}\cos \nu}},
		\label{nnb}
	\end{aligned}
\end{equation}
which decreases with $\eta$. 

\emph{ Grating lobes:} The locations of grating lobes can be obtained by letting ${\frac{{\lambda \pi \eta \cos \nu }}{{4l}}\Delta } = k\pi ,k \in {\rm{S_G}} = \left\{\pm 1, \pm 2..., \pm \left\lfloor \frac{({\underline{N}}-1)\lambda\eta \cos \nu }{4l} \right\rfloor \right\}$. Thus, the $k$-th grating lobe will appear at
\begin{equation}
	\setlength\abovedisplayskip{2pt}
	\setlength\belowdisplayskip{1pt}
	\begin{aligned}
	\Delta  = \frac{{4l}}{{\lambda \eta \cos \nu}}k,~k \in {\rm{S_G}}.
	\end{aligned}
\label{grating_lobe}
\end{equation}
Since $|\Delta|\le {\underline{N}}-1,$ it can be easily shown that only when $\eta\ge \frac{4l}{\lambda({\underline{N}}-1)\cos \nu },$ will there exist grating lobes. 
\begin{figure}[htbp]
	\setlength{\abovecaptionskip}{-0.1cm}
	\setlength{\belowcaptionskip}{-0.3cm}
	\centerline{\includegraphics[width=0.4\textwidth]{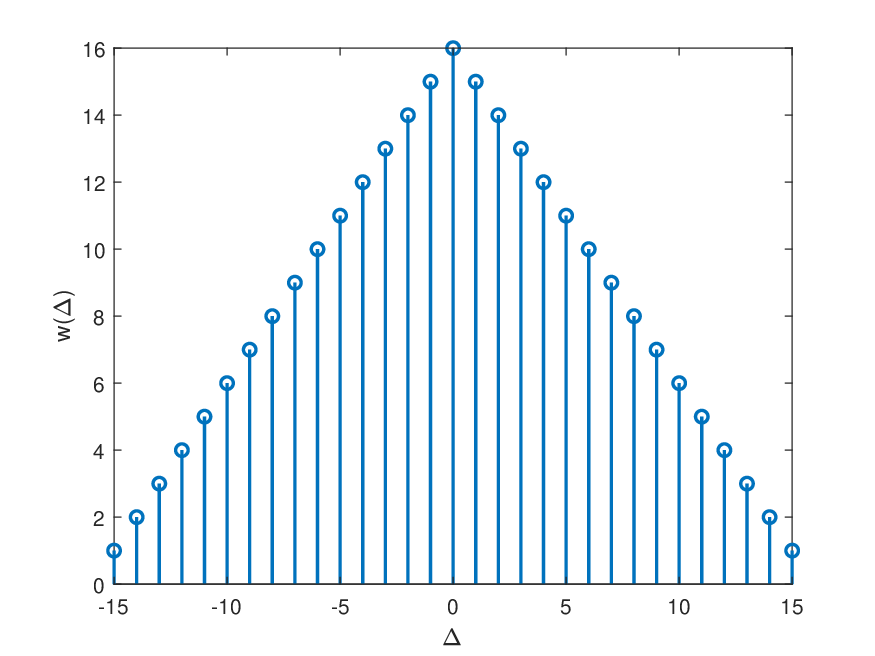}}
	\caption{Pattern of $w(\Delta )$.}
	\label{w}
\end{figure}

\begin{figure}[ht]
	\centering
	\subfigure[Pattern of $f_\eta(\Delta )$, with $\eta=22,r=10$m.]{
		\includegraphics[width=0.4\textwidth]{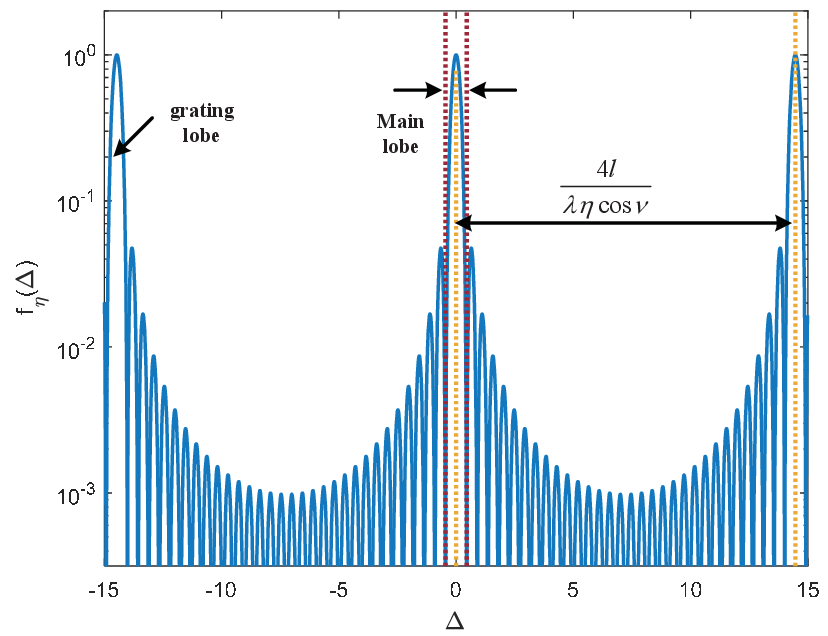}
		\label{de}
	}
	\quad
	\subfigure[Approximation of $f_\eta(\Delta )$, with $\eta=10,r=20$m, $\alpha=0.83$.]{
		\includegraphics[width=0.4\textwidth]{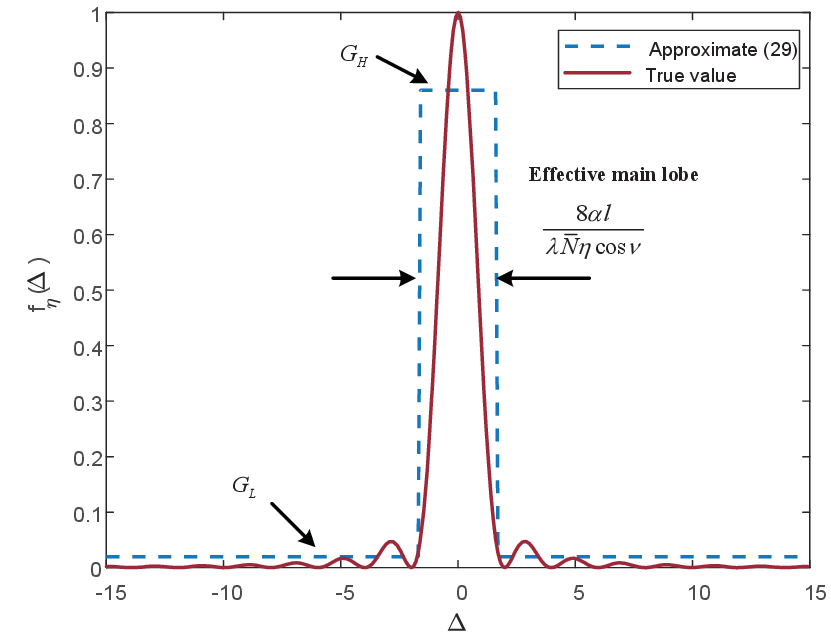}
		\label{ade}
	}
	\caption{Pattern of $f_\eta(\Delta )$. }
	\label{pattern}
\end{figure}

In order to further derive the closed-form expression of $\varepsilon$ in (\ref{closed-form}), similar to the two-lobe approximation\cite{b97}, we can use the following approximation
\begin{equation}
	\begin{aligned}
		f_\eta(\Delta) \approx \begin{cases}
			{G_H},&- t \le {\Delta} - \frac{{4lk}}{\lambda\eta \cos \nu}\le t\\
			{G_L},&{\rm{otherwise}},
		\end{cases}
		\label{twolobe}
	\end{aligned}
\end{equation}
where $2t=\frac{8\alpha l}{\lambda {\overline{N}}\eta \cos \nu}$ with $\alpha\in[0,1]$ denotes the effective beamwidth of the main lobe or grating lobe. $G_H$ and $G_L$ account for the gains of the main/grating lobe and side lobe, respectively, where $G_L\ll G_H$. Note that the values of $\alpha$, $G_H$ and $G_L$ can be obtained by performing a least-square based curve fitting. 

Therefore, ${\sum\limits_{n = 1}^{\underline{N}} {\sum\limits_{n' = 1}^{\underline{N}} {{f_\eta }(n - n')} } }$ in (\ref{closed-form}) reduces to $\sum\limits_{\Delta  \in {{\rm{S}}_{{\rm{diff}}}}} {{f_\eta }(\Delta )w(\Delta )},$ with $w(\Delta)$ being an integer valued function such that its value equals to the number of occurrences of $\Delta$ in the set ${{\rm{S}}_{{\rm{diff}}}}$\cite{b116}, as shown in Fig.~\ref{w}. Furthermore, define ${{\rm{S}}_{\rm{H}}} \buildrel \Delta \over = {\rm{ }}\left\{ {\Delta :\left| {\Delta  - \frac{{4l}}{{\lambda \eta \cos \nu }}k} \right| < t,\forall \Delta  \in {{\rm{S}}_{{\rm{diff}}}},\exists k \in {{\rm{S}}_{\rm{G}}}} \right\}{\rm{ }}$ as the set of $\Delta$ that locates within the main lobe or grating lobes. Thus, $\varepsilon$ can be further expressed as
\begin{equation}
\begin{aligned}
\varepsilon  \approx \frac{{{{\underline{N}}^2}{{\overline N}^2}}}{{{G_H}\sum\nolimits_{\Delta  \in {{\rm{S}}_{\rm{H}}}} {w(\Delta )}  + {G_L}\sum\nolimits_{\Delta  \in {{\rm{S}}_{{\rm{diff}}}},\Delta  \notin {{\rm{S}}_{\rm{H}}}} {w(\Delta )} }}.
	\end{aligned}
	\label{closed-form dof}
\end{equation}
\begin{figure*}[ht] 
	\centering
	\begin{equation}
		\setlength\abovedisplayskip{2pt}
		\setlength\belowdisplayskip{2pt}
		\begin{aligned}
			\varepsilon (\eta ) \approx \begin{cases}
				\frac{{{{\overline{N}}^2}}}{{{G_H}}},&\eta  < \frac{{4l}}{{\lambda \overline{N}(\underline{N} - 1)\cos \nu }}\\
				\frac{{{{\bar N}^2}{{\underline{N}}^2}}}{{({G_H} - {G_L})\left( { - {{\left\lfloor {\frac{{4\alpha l}}{{\lambda \bar N\eta \cos \nu }}} \right\rfloor }^2} + (2\underline{N} - 1)\left\lfloor {\frac{{4\alpha l}}{{\lambda \bar N\eta \cos \nu }}} \right\rfloor  + \underline{N}} \right) + {G_L}{{\underline{N}}^2}}},
				 &\frac{{4l}}{{\lambda \overline{N}(\underline{N} - 1)\cos \nu }} \le \eta  < \frac{{4\alpha l}}{{\lambda \overline{N}\cos \nu }}\\
				\frac{{{{\overline N}^2}\underline N}}{{{G_H} + {G_L}(\underline N - 1)}},&\eta  \ge \frac{{4\alpha l}}{{\lambda \overline{N}\cos \nu }}
			\end{cases}
			\label{final-dof}
		\end{aligned}
	\end{equation}	
	\hrulefill
\end{figure*}
\begin{figure*}[ht] 
	\centering
	\setlength{\abovecaptionskip}{-0.1cm}
	\setlength{\belowcaptionskip}{-0.3cm}
	\centerline{\includegraphics[width=0.8\textwidth]{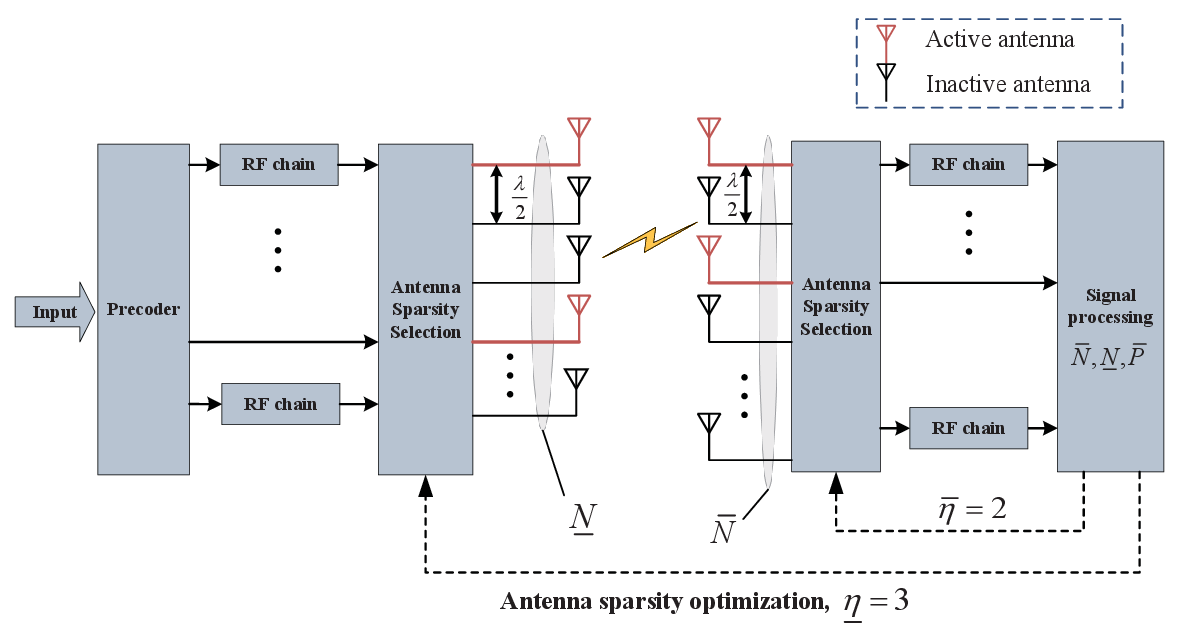}}
	\caption{Array sparsity-selection strategy for single-user MIMO communication system.}
	\label{as}
\end{figure*}
Fig.~\ref{pattern} shows the pattern of $f_\eta(\Delta )$ where $\underline{N}=16$ and $\overline{N}=32.$ In particular, Fig.~\ref{de} exhibits the main lobe and grating lobes of $f_\eta(\Delta )$. Fig.~\ref{ade} plots the two-lobe approximation of $f_\eta(\Delta )$ in (\ref{twolobe}). In order to clearly illustrate such approximation, we set small $\eta$ with no grating lobe.
\begin{theorem}\label{epsilon}
Based on the approximation in (\ref{twolobe}), the closed-form expression of the EDoF $\varepsilon$ can be expressed in terms of the sparsity parameter $\eta$, shown in (\ref{final-dof}) at the top of this page. 
\begin{IEEEproof}
	Please refer to Appendix B.
\end{IEEEproof}
\end{theorem}
Theorem \ref{epsilon} shows that the EDoF $\varepsilon$ for near-field MIMO communication is a piece-wise function, which is closely related to $\eta$, $\overline{N}$ and $\underline{N}$. 

\begin{lemma}
	From the EDoF perspective, a new boundary to separate the near-field and far-field is obtained as $l_{near-far}=\frac{{\lambda \overline{N}(\underline{N} - 1)\cos \nu }}{{4 }}\eta$. When $l > l_{near-far}$, the EDoF reduces to $\varepsilon\approx1$, which is consistent with the result for the far-field in Section \ref{far-field}.
	\begin{IEEEproof}
		According to (\ref{final-dof}), we have $\varepsilon=\frac{\overline{N}^2}{G_H}$ when $l > \frac{{\lambda \overline{N}(\underline{N} - 1)\cos \nu }}{{4 }}\eta$. Since $G_H\approx {\overline{N}^2}$, we thus obtain $\varepsilon\approx1$.
	\end{IEEEproof}
\end{lemma}
Since ${l_{{\rm{Ray}}}} \approx \frac{\lambda }{2}{\left( {\overline N\bar \eta  + \underline{N}\underline{\eta}} \right)^2} > \frac{\lambda }{2}\overline N\underline{N}\eta $, it can be easily shown that $l_{near-far}< l_{Ray}.$ Thus, a more refined partition of the near-and far-field region can be achieved.

\begin{lemma}\label{etainf}
	When $\eta  > \frac{{4\alpha l}}{{\lambda \overline{N}\cos \nu }}$, based on the fact that $G_H\approx {\overline{N}^2}\gg G_L$, the EDoF can be simplified as $\varepsilon\approx \underline{N}$.
\end{lemma}
Note that the EDoF $\varepsilon$ will reach its maximum when $\eta  > \frac{{4\alpha l}}{{\lambda \overline{N}\cos \nu }}$, which is consistent with the results in \cite{b147} by letting $\alpha=1$.

\begin{theorem}\label{antenna_selection}
	For single-user MIMO communications with the receive SNR satisfying $\bar{P}\le\frac{4}{\overline{N}\underline{N}}$, the conventional MIMO with $\eta=1$ is the optimal configuration to maximize the capacity. However, when $\bar{P}\ge\frac{4\underline{N}}{\overline{N}}$, the sparse MIMO should be used for capacity maximization. Specifically, the array sparsity $\eta\ge\frac{{4\alpha l}}{{\lambda \overline{N}\cos \nu}}$ should be chosen to achieve the maximal EDoF. For MIMO communications with $\bar P \in \left[ {\frac{4}{{\underline{N}\overline{N}}},\frac{4\underline{N}}{{\overline{N}}}} \right],$ the optimal array sparsity should lead to $\varepsilon  \approx \frac{{\sqrt {\underline{N}\overline{N}\bar P} }}{2},$ as shown in Theorem \ref{datarate}.
	\begin{IEEEproof}
	Please refer to Appendix C.
	\end{IEEEproof}
\end{theorem}
Theorem \ref{antenna_selection} implies an array sparsity-selection strategy for single-user MIMO communication system, under both far-field and near-field scenarios. For the conventional large-scale arrays with half-wavelength spacing at the BS and UE, the array sparsity can be adjusted and optimized by activating part of the antenna elements based on different channel realizations and transmission power constraints, as shown in Fig.~\ref{as}.
\begin{lemma}\label{angle}
	For a given array sparsity $\eta$, the EDoF will become larger as $|\theta|$ or $|\varphi|$ decreases.
	\begin{IEEEproof}
		According to (\ref{final-dof}), when $|\theta|$ or $|\varphi|$ decreases, $\cos\nu$ will go large. Therefore, based on the analysis in Appendix C, Lemma \ref{angle} can be easily obtained.
	\end{IEEEproof}
\end{lemma}
Note that Lemma \ref{angle} is reasonable since the effective aperture can be enlarged by placing the transceiver parallel to each other.

To sum up, for the single-user communication in the near-field region, the DoF of MIMO channel can be improved by deploying the sparse MIMO when the transmit power is sufficiently large. Therefore, the spatial multiplexing gain and data rate can be enhanced, compared with conventional MIMO.
\section{Multi-user Sparse MIMO}
In this section, the LoS-dominant multi-user scenario is considered for both far-field and near-field sparse MIMO communications.
\subsection{Far-field multi-user Sparse MIMO}\label{far-field multi-user}
We first consider the far-field scenario, where the LoS dominant channel matrix ${\bf H}_k$ can be expressed as
\begin{equation}
		\setlength\abovedisplayskip{2pt}
	\setlength\belowdisplayskip{1pt}
	\begin{aligned}
	{{\bf{H}}_k} = {\beta _k}{\bf{b}}(\varphi_k){{\bf{a}}_k^H}(\theta_k),
	\end{aligned}
	\label{h-far}
\end{equation}
where ${\bf{b}}(\varphi)$ and ${\bf{a}}_k(\theta)$ are the receive and transmit steering vectors, which can be modelled based on (\ref{steer vector}) and (\ref{r_far}) as
\begin{equation}
		\setlength\abovedisplayskip{2pt}
	\setlength\belowdisplayskip{1pt}
	\begin{aligned}
	{\bf{b}}(\varphi ) &= {[{e^{ - j\pi \left( { - \frac{{{N_{\rm{BS}}} - 1}}{2}} \right){\eta _{{\rm{BS}}}}\sin \varphi }},...,{e^{ - j\pi \left( {\frac{{{N_{\rm{BS}}} - 1}}{2}} \right){\eta _{{\rm{BS}}}}\sin \varphi }}]^T},\\
	{\bf{a}}_k(\theta ) &= {[{e^{ - j\pi \left( { - \frac{{{N_{\rm{UE}}} - 1}}{2}} \right){\eta _{{\rm{UE}}}}\sin \theta }},...,{e^{ - j\pi \left( {\frac{{{N_{\rm{UE}}} - 1}}{2}} \right){\eta _{{\rm{UE}}}}\sin \theta }}]^T}.
	\end{aligned}
\end{equation}

It is obvious from (\ref{h-far}) that for far-field MIMO communications, the rank of the channel always equals to $1$. Therefore, one data stream can be transmitted for each UE, and the transmitted signal of $k$ should be expressed as ${\bf x}_k = \sqrt{P_k}{\bf t}_ks_k$, where ${\bf t}_k \in \mathbb{C}^{{N_{\rm{UE}}}\times 1}$ is the unit norm transmit beamforming vector, and $s_k$ and $P_k$ denote the symbol and transmit power of UE $k$, respectively. 

To detect the signal for UE $i$, a linear receive beamforming vector$\ {{\bf{v}}_i}\ \in \mathbb{C}^{{N_{\rm{BS}}}\times 1}$ with $||{{\bf{v}}_i}||=1$ can be used and the resulting signal is given by
\begin{equation}
	\setlength\abovedisplayskip{2pt}
	\setlength\belowdisplayskip{2pt}
	\begin{aligned}
		{y_i} &= {\bf{v}}_i^H{\bf{y}} \\
		&= {\bf{v}}_i^H{{\bf{H}}_i}\sqrt {{P_i}} {{\bf{t}}_i}{s_i} + {\bf{v}}_i^H\sum\limits_{k = 1,k \ne i}^K {{{\bf{H}}_k}\sqrt {{P_k}} {{\bf{t}}_k}{s_k}}  + {\bf{v}}_i^H{\bf{n}}.
	\end{aligned}
	\label{bf}
\end{equation}
Thus, the signal-to-interference-plus-noise ratio (SINR) for UE $i$ is expressed as
\begin{equation}
	\setlength\abovedisplayskip{2pt}
	\setlength\belowdisplayskip{1pt}
	\begin{aligned}
		{\gamma _i} &= \frac{{{P_i}{{\left| {{\bf{v}}_i^H{{\bf{H}}_i}{{\bf{t}}_i}} \right|}^2}}}{{\sum\limits_{k = 1,k \ne i}^K {{P_k}{{\left| {{\bf{v}}_i^H{{\bf{H}}_k}{{\bf{t}}_k}} \right|}^2}}  + {\sigma ^2}}}\\
		& = \frac{{{{\bar P}_i}{{\left| {{\bf{v}}_i^H{\bf{b}}({\varphi _i})} \right|}^2}{{\left| {{{\bf{a}}_i^H}({\theta _i}){{\bf{t}}_i}} \right|}^2}}}{{\sum\limits_{k = 1,k \ne i}^K {{{\bar P}_k}{{\left| {{\bf{v}}_i^H{\bf{b}}({\varphi _k})} \right|}^2}{{\left| {{{\bf{a}}_k^H}({\theta _k}){{\bf{t}}_k}} \right|}^2}}  + 1}},i = 1,...,K
		\label{sinr}
	\end{aligned}
\end{equation}
where $\bar{P_i} \triangleq \frac{{{{\left| {{\beta _i}} \right|}^2}{P_i}}}{{{\sigma ^2}}},$ and the achievable sum rate in bits/second/Hz is 
\begin{equation}
	\setlength\abovedisplayskip{-1pt}
	\setlength\belowdisplayskip{1pt}
	\begin{aligned}
R = \sum\limits_{i = 1}^K {{{\log }_2}(1 + {\gamma _i})} .
		\label{sumrate}
	\end{aligned}
\end{equation}

\begin{figure}[htbp]
	\setlength{\abovecaptionskip}{-0.1cm}
	\setlength{\belowcaptionskip}{0.1cm}
	\centerline{\includegraphics[width=0.4\textwidth]{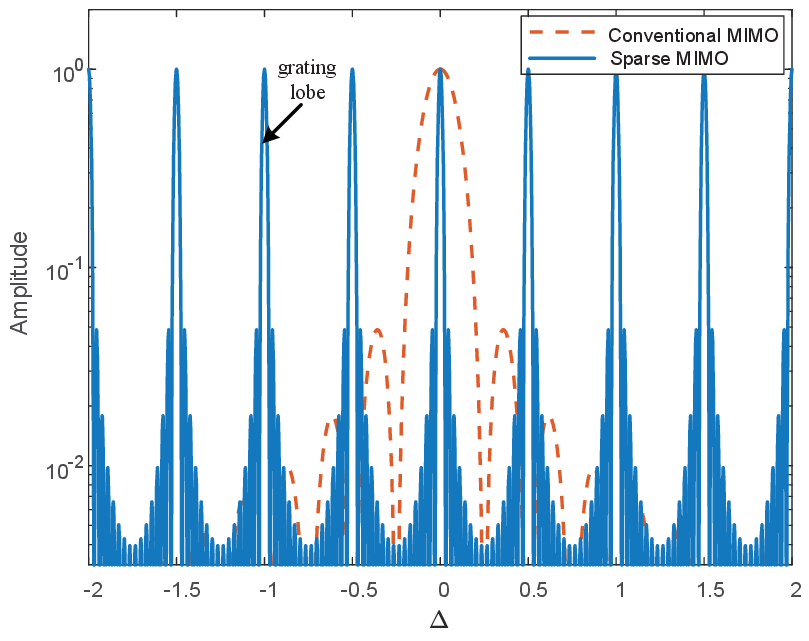}}
	\caption{Beam pattern of sparse and conventional MIMO ($\eta_{\rm{BS}}=4$)\cite{b109}.}
	\label{patterns}
\end{figure}

By using the maximum-ratio transmission (MRT) and maximal-ratio combining (MRC) beamforming at the UE and BS, respectively, i.e., ${{\bf{t}}_i} = \frac{{{\bf{a}}(\theta _i)}}{{\left\| {{\bf{a}}(\theta _i)} \right\|}},{{\bf{v}}_i} = \frac{{{\bf{b}}(\varphi _i)}}{{\left\| {{\bf{b}}(\varphi _i)} \right\|}},$ the SINR in (\ref{sinr}) becomes
\begin{equation}
	\setlength\abovedisplayskip{2pt}
	\setlength\belowdisplayskip{1pt}
	\begin{aligned}
		{\gamma _i} = \frac{{{{\bar P}_i}{N_{\rm{UE}}}{N_{\rm{BS}}}}}{{{N_{\rm{UE}}}{N_{\rm{BS}}}\sum\limits_{i = 1,i \ne k}^K {{{\bar P}_k}{{\left| {\frac{1}{{N_{\rm{BS}}}}{{\bf{b}}^H}(\varphi _i){\bf{b}}(\varphi_k)} \right|}^2}}  + 1}}.
		\label{sinr-after}
	\end{aligned}
\end{equation}
It can be observed from (\ref{sinr-after}) that with the MRT and MRC beamforming, the performance of multi-user MIMO system only depends on the sparsity of the receive array at the BS, but irrespective of that of the transmit array at the UE side due to MRT-based beamforming. Furthermore, we define the beam pattern as
\begin{equation}
	\setlength\abovedisplayskip{2pt}
	\setlength\belowdisplayskip{1pt}
\small
	\hspace{-1ex}
	\begin{aligned}
		G_\eta(\Delta_{ik})\triangleq{\left| {\frac{1}{{N_{\rm{BS}}}}\sum\limits_{n = 0}^{{N_{\rm{BS}}} - 1} {{b}_n^*({\varphi _i}){{b}_n}({\varphi _k})} } \right|^2} = {\left| {\frac{{\sin \left( {\frac{\pi }{2}{N_{\rm{BS}}}\eta_{BS} {\Delta _{ik}}} \right)}}{{{N_{\rm{BS}}}\sin \left( {\frac{\pi }{2}\eta_{BS} {\Delta _{ik}}} \right)}}} \right|^2},
		\label{rho}
	\end{aligned}
\end{equation}
where ${\Delta _{ik}} \triangleq \sin {\varphi _k} - \sin {\varphi _i}\in [-2, 2]$ denotes the spatial angle difference. Indeed, the SINR of each UE $k$ significantly depends on the beam pattern $G_\eta(\Delta_{ik})$, formulated as a function of the spatial angle difference $\Delta_{ik}$ with the receive array sparsity $\eta_{\rm{BS}}$. When the sparse array is deployed at the BS, the main lobe of $G_\eta(\Delta_{ik})$ tends to become narrower, which yields smaller IUI for two UEs with small spatial angle difference $\Delta_{ik}$. However, undesired grating lobes that have equal amplitude and width as the main lobe will also be generated, leading to severer IUI when two UEs lie in the grating lobe of each other, as shown in Fig.~\ref{patterns}. 
\begin{figure}[htbp]
	\setlength{\abovecaptionskip}{-0.1cm}
	\setlength{\belowcaptionskip}{0.1cm}
	\centerline{\includegraphics[width=0.4\textwidth]{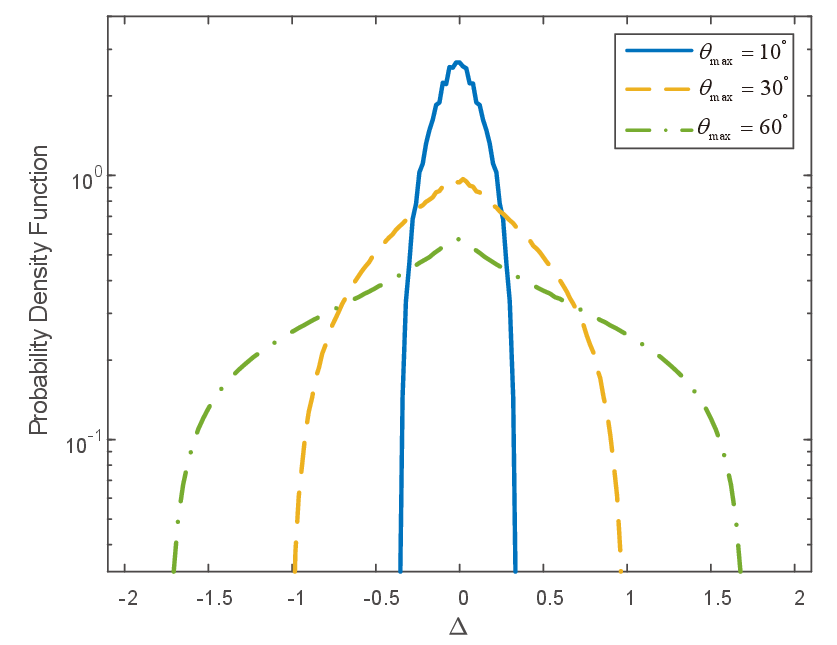}}
	\caption{PDF of spatial angle difference $\Delta$\cite{b109}.}
	\label{delta}
\end{figure}
\begin{figure*}[ht] 
	\centering
	\begin{equation}
		\setlength\abovedisplayskip{2pt}
		\setlength\belowdisplayskip{2pt}
		\begin{aligned}
			{p(\eta_{{\rm{BS}}})} \approx 
			\begin{cases}
				{\frac{{\alpha \left( {\left( {2{n_{\max }} + 1} \right)\sin {\varphi _{\max }} - \frac{\alpha }{{4N_{\rm{BS}}\eta_{\rm{BS}} }} - \frac{{{n_{\max }}\left( {{n_{\max }} + 1} \right)}}{\eta_{\rm{BS}} }} \right)}}{{{{\sin }^2}{\varphi _{\max }}N_{\rm{BS}}\eta_{\rm{BS}} }}},&\sin\varphi_{\max}\ge {{\frac{\alpha }{{2N_{\rm{BS}}\eta_{\rm{BS}} }}}} \\
				1,&\sin {\varphi _{\max}}<{{\frac{\alpha }{{2N_{\rm{BS}}\eta_{\rm{BS}} }}}} 
			\end{cases}.
			\label{p}
		\end{aligned}
	\end{equation}	
	\hrulefill
\end{figure*}

Fig.~\ref{delta} plots the distribution of $\Delta$ when the AoAs of the UEs are uniformly distributed in $\theta_k\in [-\theta_{\max},\theta_{\max}], \forall k$\cite{b109}. It is observed that the probability of $\Delta$ decreases with the increase of $|\Delta|$. This indicates that the non-uniform distribution of $\Delta$ may act as a natural filter, to automatically reject higher-order grating lobes of sparse MIMO. As a result, compared to conventional MIMO, it becomes less likely that $\Delta$ will fall into either the main lobe or grating lobe. Furthermore, since sparse arrays have much narrower main lobe than the compact counterpart, they are less likely to suffer from severe IUI and hence a better performance is expected. 

Furthermore, the two-lobe approximation can be used to approximate $G_\eta(\Delta_{ik})$. Assume the receive SNR ${{\bar P}_k}={{\bar P}},\forall k,$ and the location of UE $k$, i.e., $X_k\triangleq\sin\varphi_k$ follows i.i.d. distribution with the support $[-\sin\varphi_{\max},\sin\varphi_{\max}]$ over $k$. Therefore, the cumulative distribution function (CDF) of the achievable data rate in (\ref{sumrate}) can be expressed as \cite{b109}
	\begin{equation}
		\setlength\abovedisplayskip{2pt}
		\setlength\belowdisplayskip{2pt}
		\begin{aligned}
			\begin{array}{l}
				F = \mathrm{Pr}(R\leq \bar R)= 1 - \sum\limits_{q = 0}^T {C_{K - 1}^q}{{{{p}} }^q}{{(1 - {p})}^{(K - 1) - q}},
			\end{array}
		\end{aligned}
		\label{bi}
	\end{equation}
where $C_{K - 1}^q = \frac{{(K - 1)!}}{{q!(K - 1 - q)!}}$ accounts for the binomial coefficient. $T = \left\lfloor {\frac{Y-(K-1)G_s}{{G_m - G_s}}} \right\rfloor,$ and $Y = \frac{1}{{{2^{\bar R}} - 1}} - \frac{1}{{\bar PN}}$. $p(\eta)$ denotes the probability that two UEs lie in the main lobe or the grating lobe for a given sparsity $\eta_{\rm{BS}}$, which can be expressed as (\ref{p}) at the top of this page~\cite{b109}, where ${n_{\max}} = \left\lfloor {\eta_{\rm{BS}}\sin\varphi_{\max}  - \frac{{\alpha}}{2N_{\rm{BS}}}} \right\rfloor, $ and $\alpha\in[0,1]$ is the same as that in Section \ref{near-single}. Note that the probability $p$ for conventional MIMO can be easily expressed as $p(1)$, by letting $\eta_{{\rm{BS}}}=1$ in (\ref{p}).

Furthermore, by comparing $p(\eta_{{\rm{BS}}})$ with $p(1)$, it can be found that for $\sin {\varphi _{\max }} \in \left( \sin\underline{\varphi},\sin\overline{\varphi}\right)$, we always have $p(\eta_{{\rm{BS}}})<p(1)$, where $\sin\underline{\theta}\triangleq \frac{\alpha }{{2N_{{\rm{BS}}}\eta_{{\rm{BS}}} }}$ and $\sin\overline{\theta}\triangleq	\frac{{\eta_{{\rm{BS}}}  + \sqrt {{\eta_{{\rm{BS}}} ^2} - \frac{\alpha }{N_{{\rm{BS}}}}({\eta_{{\rm{BS}}} ^2} + 1 - \frac{\alpha }{N_{{\rm{BS}}}})} }}{{2\eta_{{\rm{BS}}} }}.$ This demonstrates that for a relatively wide range of the user distribution, sparse MIMO strictly outperforms conventional MIMO from the IUI mitigation perspective.

\subsection{Near-field multi-user Sparse MIMO}
Similar to (\ref{def}), for multi-user MIMO system, the data rate of UE $k$ can be expressed as
\begin{equation}
	\setlength\abovedisplayskip{2pt}
	\setlength\belowdisplayskip{2pt}
	\begin{aligned}
	{R_k} = {\log _2}\det &\left[ {{{\bf{I}}_{{N_{{\rm{BS}}}}}} + {{\bf{H}}_k}{{\bf{Q}}_k}{\bf{H}}_k^H} \right. \\
		&\left.\times {\left(\sum\limits_{i \ne k} {{{\bf{H}}_i}{{\bf{Q}}_i}{\bf{H}}_i^H + {\sigma ^2}{{\bf{I}}_{{N_{{\rm{BS}}}}}}} \right)^{ - 1}}\right],
	\end{aligned}
	\label{near_multi_rate}
\end{equation}
and the sum rate is given by
\begin{equation}
	\setlength\abovedisplayskip{2pt}
	\setlength\belowdisplayskip{2pt}
	\begin{aligned}
		{R_{sum}} = \sum\limits_{k = 1}^K {{R_k}}.
	\end{aligned}
	\label{near_sumrate}
\end{equation}
By considering the case without CSIT, i.e., ${\bf Q}_k = \frac{P_k}{{N_{\rm{UE}}}}{\bf I}_{N_{\rm{UE}}}$, and assuming $P_k=P,\forall k$, (\ref{near_multi_rate}) can be further simplified as
\begin{equation}
	\setlength\abovedisplayskip{2pt}
	\setlength\belowdisplayskip{2pt}
	\begin{aligned}
{R_k} = {\log _2}\det &\left[ {{\bf{I}}_{{N_{{\rm{BS}}}}}} + {{\bf{H}}_k}{\bf{H}}_k^H \right.\\
	&\left.\times {{\left( {\sum\limits_{i \ne k} {{{\bf{H}}_i}{\bf{H}}_i^H + \frac{{{\sigma ^2}{N_{{\rm{UE}}}}}}{P}{{\bf{I}}_{{N_{{\rm{BS}}}}}}} } \right)}^{ - 1}} \right].
	\end{aligned}
	\label{near_multi_rate_equal}
\end{equation}

However, it is challenging to obtain the closed-form expression of the data rate in (\ref{near_multi_rate_equal}). Therefore, we will evaluate the performance of near-field multi-user MIMO communications numerically.
\section{Numerical results}
In this section, numerical results are provided to show the performance of single- and multi-user sparse MIMO. 
\subsection{Single-user Sparse MIMO}
\begin{figure}[htbp]
	\setlength{\abovecaptionskip}{-0.1cm}
	\setlength{\belowcaptionskip}{-0.3cm}
	\centerline{\includegraphics[width=0.5\textwidth]{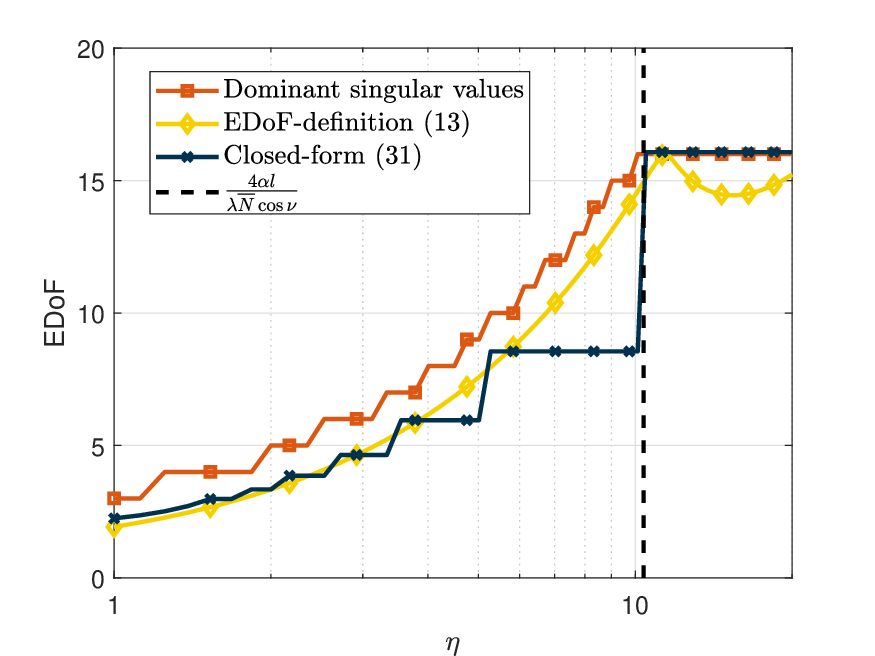}}
	\caption{EDoF $\varepsilon$ versus array sparsity for the single-user case.}
	\label{edof}
\end{figure}
\begin{figure*}[hb]
	\centering
	\subfigure[$\bar{P}=-30$dB]{
		\includegraphics[width=0.3\textwidth]{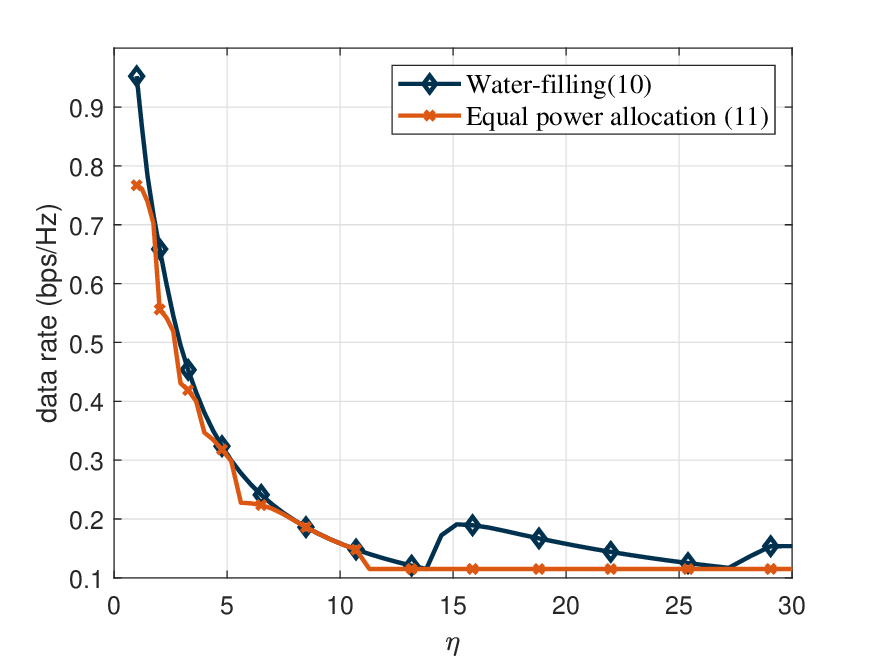}
	}
	\quad
	\subfigure[$\bar{P}=-10$dB]{
		\includegraphics[width=0.3\textwidth]{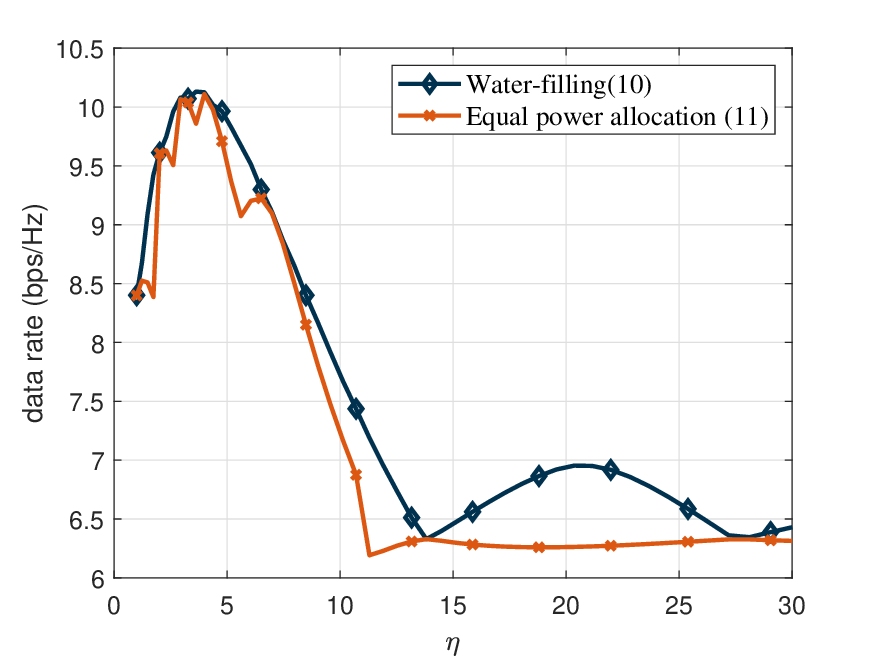}
	}
	\quad
	\subfigure[$\bar{P}=10$dB]{
		\includegraphics[width=0.3\textwidth]{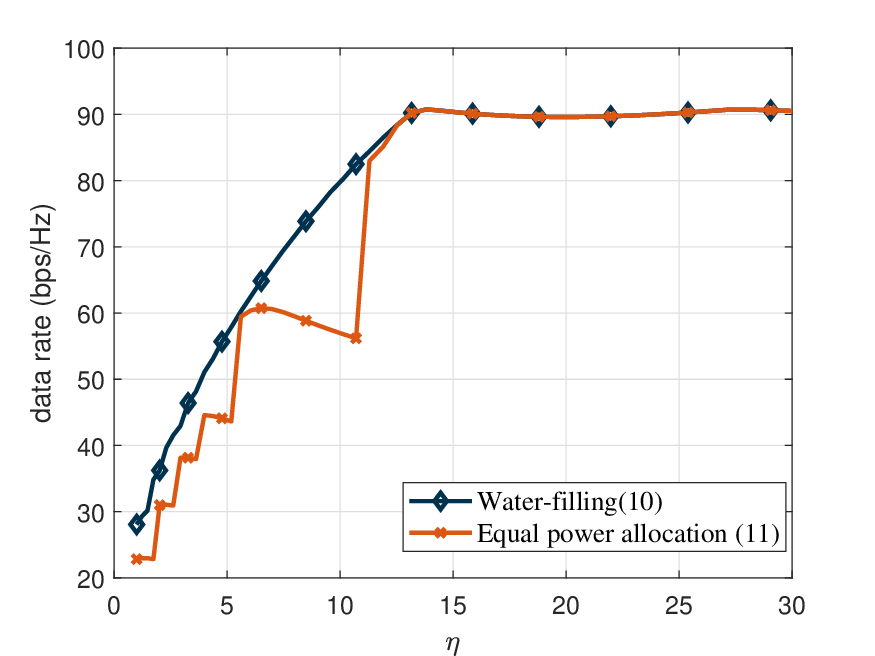}
	}
	\caption{Data rate versus $\eta$ under different receive SNR. }
	\label{data}
\end{figure*}
For the setup with $N_{\rm{UE}}=16$, $N_{\rm{BS}}=128$, and $l=40$m (i.e., near-field), Fig.~\ref{edof} plots the derived EDoF $\varepsilon$ in (\ref{final-dof}), legend as ``Closed-form (\ref{final-dof})". We also plot the DoF based on dominant singular values (``Dominant singular values") and its original definition in (\ref{edof_Def}) for comparison. The values of $\alpha$ and $G_H$ are obtained via curve fitting, given by $\alpha=0.91$ and ${G_H} = 0.5N_{\rm{BS}}^2$. The dashed line shows the optimal array sparsity in Lemma \ref{etainf}. It can be observed that both the exact and derived closed-form expressions of EDoF increase with $\eta$ before reaching their maxima, which indicates the advantages of deploying the sparse MIMO. Furthermore, when $\eta>\frac{{4\alpha l}}{{\lambda (N_{\rm{BS}} - 1)\cos \nu }}$, the EDoF will approach to a constant. This is consistent with the analysis in Theorem \ref{antenna_selection}, and provides a theoretical threshold when the EDoF can achieve its maxima. Note that the error between (\ref{final-dof}) and (\ref{edof_Def}) around $\eta=\frac{{4\alpha l}}{{\lambda (N_{\rm{BS}} - 1)\cos \nu }}$ is due to the ``$\left\lfloor  \cdot  \right\rfloor $" operation.

Fig.~\ref{data} shows the achievable data rate versus $\eta$ under different receive SNR $\bar{P}$. The achievable data rate using the optimal water-filling power allocation is also plotted for comparison. It can be observed that the general trend for each SNR region is consistent with the analysis in Theorem~\ref{datarate}. Besides, the derived data rate provides a lower bound of the optimal water-filling in all SNR regimes.
\subsection{Multi-user Sparse MIMO}
Unless otherwise stated, we set ${N_{\rm{UE}}}=8,$ and $K=20$. All the UEs are uniformly distributed as $\varphi_k \in [-\varphi_{\rm{max}},\varphi_{\rm{max}}]$ with identical distance $l$. The transmit SNR is set to be $\frac{P}{\sigma^2} = 90$dB. For the multi-path channel in (\ref{channel}), we use the ``one-ring" model, with $Q_k=5$ and $R=3$m denoting the number of multi-paths and radius of each ring, respectively\cite{b88}. Besides, we set $F_k=20$dB, $\forall k$. 

\begin{figure}[htbp]
	\setlength{\abovecaptionskip}{-0.1cm}
	\setlength{\belowcaptionskip}{-0.3cm}
	\centerline{\includegraphics[width=0.5\textwidth]{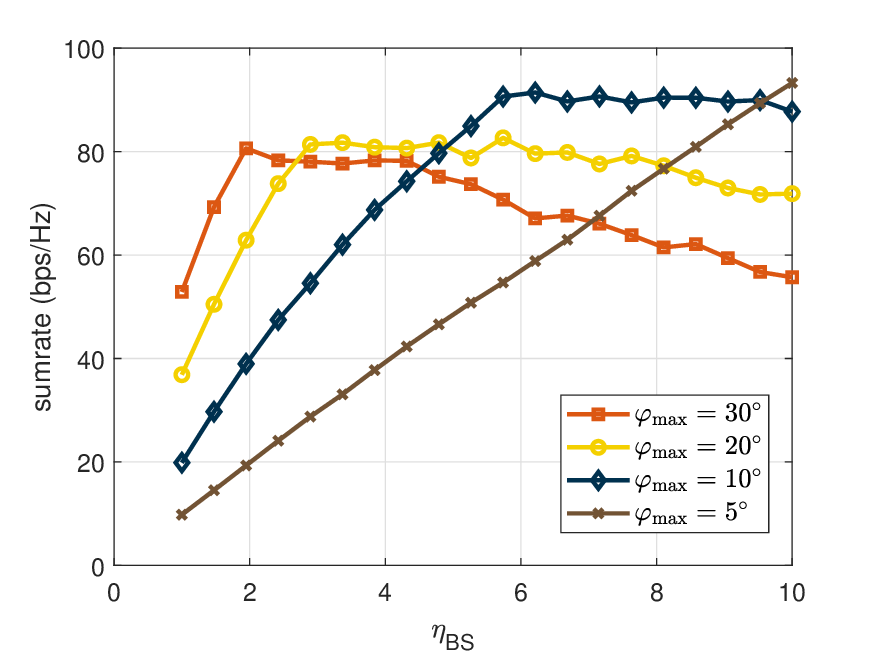}}
	\caption{Far-field sum rate versus BS array sparsity ($\eta_{{\rm{UE}}}=1$).}
	\label{multiuser_BS}
\end{figure}

\begin{figure}[htbp]
	\setlength{\abovecaptionskip}{-0.1cm}
	\setlength{\belowcaptionskip}{-0.3cm}
	\centerline{\includegraphics[width=0.5\textwidth]{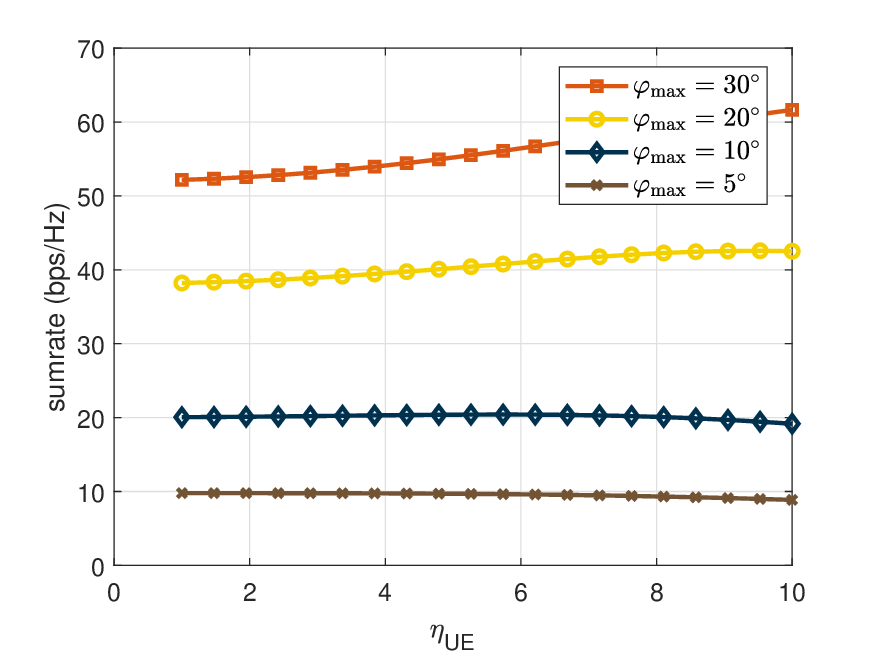}}
	\caption{Far-field sum rate versus UE array sparsity ($\eta_{{\rm{BS}}}=1$).}
	\label{multiuser_UE}
\end{figure}

Fig.~\ref{multiuser_BS} and Fig.~\ref{multiuser_UE} show the achievable sum rate in (\ref{sumrate}) versus the array sparsity at the BS or the UE with varying $\varphi_{\max}$. We set ${N_{\rm{BS}}}=64$, and $l$ is $120$m (i.e., far-field). It can be seen that when $\eta_{{\rm{UE}}}=1$, the sum rate first increases with $\eta_{{\rm{BS}}}$ and then decreases, indicating the benefit of using sparse array at the BS and the importance of choosing $\eta_{{\rm{BS}}}$ properly. Besides, for relatively small $\varphi_{\max}$, the enhancement of sum rate becomes more obvious by increasing $\eta_{{\rm{BS}}}$. This is reasonable since larger spatial resolution and better interference mitigation ability are needed for closely located UEs, and thus larger receive array sparsity should be chosen. On the other hand, when $\eta_{{\rm{BS}}}$ is fixed, the sum rate will not increase as $\eta_{{\rm{UE}}}$ increases, which validates our analysis in Section~\ref{far-field multi-user}.

\begin{figure}[htbp]
	\setlength{\abovecaptionskip}{-0.1cm}
	\setlength{\belowcaptionskip}{-0.3cm}
	\centerline{\includegraphics[width=0.5\textwidth]{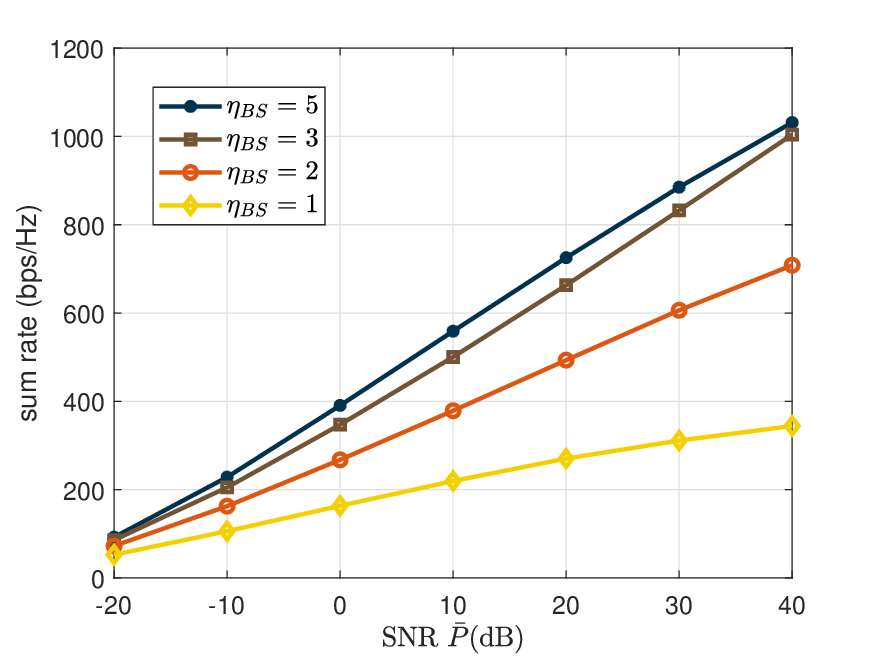}}
	\caption{Near-field sum rate versus receive SNR ($\eta_{{\rm{UE}}}=1$).}
	\label{multiuser_SNR}
\end{figure}
\begin{figure}[htbp]
	\setlength{\abovecaptionskip}{-0.1cm}
	\setlength{\belowcaptionskip}{-0.3cm}
	\centerline{\includegraphics[width=0.5\textwidth]{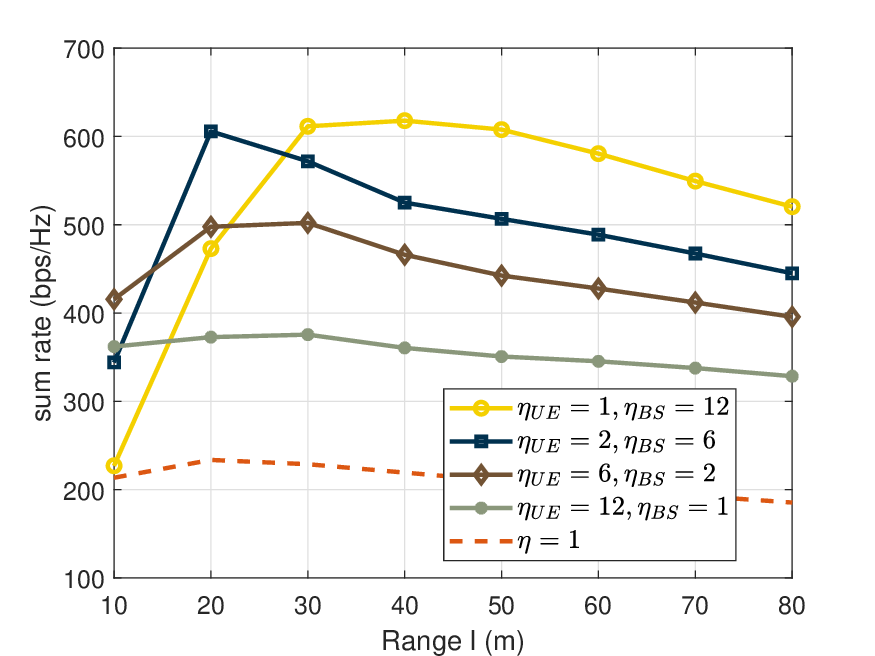}}
	\caption{Sum rate versus user range $l$.}
	\label{multiuser_eta}
\end{figure}
Fig.~\ref{multiuser_SNR} plots the sum rate versus the receive SNR $\bar{P}$ for different $\eta_{{\rm{BS}}}$. In this case, we set ${N_{\rm{BS}}}=128$, $l=40$m, (i.e., near-field) and $\varphi_{\max}=20^\circ$. It can be observed that for all the four considered cases, the sum rates increase as the receive SNR grows. Furthermore, for a given $\bar{P}$, the sum rates tend to increase drastically with larger array sparsity. However, when $\eta_{{\rm{BS}}}$ is large, e.g., $\eta_{{\rm{BS}}}>5$, the improvement of sum rates become less evident. This is reasonable since the DoF has reached its maximum, so it cannot be further enlarged by increasing the array sparsity.

Furthermore, Fig.~\ref{multiuser_eta} plots the achievable sum rate versus the user range $l$, where the array sparsity is fixed as $\eta=\eta_\mathrm{BS}\eta_\mathrm{UE}=12$ and $\theta_k=0$. Conventional MIMO with $(\eta_{\rm{UE}}, \eta_{\rm{BS}})=(1,1)$ is also plotted for comparison. In order to analyze the influence of array sparsity at the UE and BS, respectively, four cases are considered, namely $(\eta_{\rm{UE}}, \eta_{\rm{BS}})=(1,10),~(2, 5),~(4, 2.5),$ or $(5, 2).$

It can be seen that for all the four cases considered, sparse MIMO can achieve much larger sum sate, compared with the conventional MIMO with $\eta=1$, and the achievable sum rate decreases for large $l$ because of higher path loss. Besides, when $l>30$m, deploying sparse array at the BS can achieve better performance than sparse array at the UE, which is consistent with the far-field analysis in Section~\ref{far-field multi-user}. However, for relatively small distance, e.g., $l<30$m, deploying sparse array at UE can achieve higher sum rate. However, when further increasing the array sparsity at the UE, while BS is equipped with compact array, e.g., $(\eta_{\rm{UE}}, \eta_{\rm{BS}})=(12,1)$, the sum rate deteriorates in general, which indicates the importance of choosing proper sparsity for BS and UE under different range. 
\section{CONCLUSION}
In this paper, we investigated the performance of uniform sparse MIMO for wireless communication systems. We first studied the spatial multiplexing gain of single-user sparse MIMO communication system, and derived the closed-form expression of the near-field EDoF, which was shown to increase with larger array sparsity before reaching its upper bound. By analyzing the scaling law for achievable data rate, we also proposed an array sparsity-selection strategy. Furthermore, we studied the multi-user sparse MIMO communication, and our results demonstrated that deploying sparse MIMO is beneficial for the IUI mitigation. Numerical results were provided to corroborate our theoretical analysis.
\section*{APPENDIX A: Proof of Theorem \ref{datarate}}
Based on (\ref{single_rate_x}), the derivative of $R(x)$ with respect to $x$ can be expressed as
 \begin{equation}
	\setlength\abovedisplayskip{2pt}
	\setlength\belowdisplayskip{2pt}
	\begin{aligned}
R'(x) = \frac{{\partial R(x)}}{{\partial x}} = \frac{{\frac{{2C{x^2}}}{{1 + C{x^2}}} - \ln \left( {1 + C{x^2}} \right)}}{{{x^2}\ln 2}}.
	\end{aligned}
	\label{a2}
\end{equation}
By letting $R'(x)=0,$ we have the following equation
\begin{equation}
	\setlength\abovedisplayskip{2pt}
	\setlength\belowdisplayskip{2pt}
	\begin{aligned}
\frac{{2C{x^2}}}{{1 + C{x^2}}} = \ln (1 + C{x^2}).
	\end{aligned}
	\label{eq}
\end{equation}
With $x>0$, the unique solution for the equation (\ref{eq}) is
 \begin{equation}
	\setlength\abovedisplayskip{2pt}
	\setlength\belowdisplayskip{2pt}
	\begin{aligned}
x_\mathrm{opt}=\sqrt{\frac{e^{W(-\frac{2}{e^2})+2}-1}{C}},
	\end{aligned}
	\label{a3}
\end{equation}
where $W(\cdot)$ is the inverse of the function $f(W) = We^W,$ and we obtain $x_\mathrm{opt}\approx\frac{2}{{\sqrt C }}$ by using numerical analysis. 

Furthermore, since the function $R'(x)$ is continuous and $R'(\frac{1}{{\sqrt C }}) = \frac{{C\left( {1 - \ln 2} \right)}}{{\ln 2}} > 0$, we always have $R'(x)>0$ when $x\in(0,x_\mathrm{opt})$, which means $R(x)$ increases monotonically for $x\in(0,x_\mathrm{opt})$. 

Similarly, by letting $x=\frac{3}{\sqrt{C}}>x_\mathrm{opt}$, we have $R'(\frac{3}{{\sqrt C }}) = \frac{{C\left( {1.8 - \ln 10} \right)}}{{9\ln 2}} \approx  - 0.08C < 0$. Therefore, when $x>x_\mathrm{opt}$, we always have $R'(x)<0$, which means $R(x)$ will decrease monotonically. Therefore, $x_\mathrm{opt}$ achieves the maximum value of $R(x)$.

\section*{APPENDIX B: Proof of Theorem \ref{epsilon}}
Based on (\ref{closed-form dof}), we first define
\begin{equation}
	\setlength\abovedisplayskip{2pt}
\setlength\belowdisplayskip{2pt}
\begin{aligned}
		S &\triangleq {G_H}\sum\nolimits_{\Delta  \in {{\rm{S}}_{\rm{H}}}} {w(\Delta )}  + {G_L}\sum\nolimits_{\Delta  \in {{\rm{S}}_{{\rm{diff}}}},\Delta  \notin {{\rm{S}}_{\rm{H}}}} {w(\Delta )} \\
		&= {G_H}\sum\nolimits_{\Delta  \in {{\rm{S}}_{\rm{H}}}} {w(\Delta )}  + {G_L}\left( {{\underline N^2} - \sum\nolimits_{\Delta  \in {{\rm{S}}_{\rm{H}}}} {w(\Delta )} } \right)\\
		&= {G_L}{\underline N^2} + ({G_H} - {G_L})\sum\nolimits_{\Delta  \in {{\rm{S}}_{\rm{H}}}} {w(\Delta )},
\end{aligned}
\label{S}
\end{equation}
where ${{\rm{S}}_{\rm{H}}} = \left\{ {\Delta \left| {\left| {\Delta  - \frac{{4l}}{{\lambda \eta \cos \nu }}k} \right| < t,\forall \Delta  \in {{\rm{S}}_{{\rm{diff}}}},\exists k \in {{\rm{S}}_{\rm{G}}}} \right.} \right\}.$ It can be seen that the expression of $S$ in (\ref{S}) is highly related to $\sum\nolimits_{\Delta  \in {{\rm{S}}_{\rm{H}}}} {w(\Delta )}.$ Therefore, we analyze $\sum\nolimits_{\Delta  \in {{\rm{S}}_{\rm{H}}}} {w(\Delta )} $ for the main lobe and grating lobes, denoted as $S_0$ and $S_+$, respectively. 

Based on (\ref{grating_lobe}), when $\frac{{( \underline{N}- 1)\lambda \eta \cos \nu }}{{4l}} < 1$,  i.e., $\eta  < \frac{{4l}}{{(\underline{N}- 1)\lambda \cos \nu }},$ no grating lobe will be generated. In this case, we have $k=0$ and ${{\rm{S}}_{\rm{H}}} = \left\{ {\Delta \left| {\left| \Delta  \right| < t,\forall \Delta  \in {{\rm{S}}_{{\rm{diff}}}}} \right.} \right\}$, thus yielding
\begin{equation}
	\setlength\abovedisplayskip{2pt}
\setlength\belowdisplayskip{2pt}
	\begin{aligned}
  S_0 =&  {\underline N + 2\sum\limits_{\delta  = 1}^{\left\lfloor t \right\rfloor } {\left( {\underline N - \delta } \right)} }  \\
		=&\underline N + 2{G_H}\underline N\left\lfloor t \right\rfloor  - {G_H}\left\lfloor t \right\rfloor \left( {\left\lfloor t \right\rfloor  + 1} \right)\\
		=&\left( { - {{\left\lfloor t \right\rfloor }^2} + (2\underline N - 1)\left\lfloor t \right\rfloor  + \underline N} \right).
		\label{int_0}
	\end{aligned}
\end{equation}
Thus, for $\eta  < \frac{{4l}}{{(\underline{N}- 1)\lambda \cos \nu }},$ $S$ in (\ref{S}) reduces to 
\begin{equation}
	\begin{aligned}
		S={G_L}{\underline N^2} + ({G_H} - {G_L})S_0.
	\end{aligned}
\label{nograting}
\end{equation}

On the other hand, for $\eta  \ge \frac{{4l}}{{(\underline{N}- 1)\lambda \cos \nu }}$, grating lobes need to be considered. Based on the symmetry property of $k$, we first consider $k> 0$, and we have ${{\rm{S}}_{\rm{H}}} = \left\{ {\Delta \left| {\left| {\Delta  - \frac{{4lk}}{{\lambda \eta \cos \nu }}} \right| < t,\forall \Delta  \in {{\rm{S}}_{{\rm{diff}}}},k \in {{\rm{S}}_{\rm{G}}},k > 0} \right.} \right\}$. In this case, we have 
\begin{equation}
	\setlength\abovedisplayskip{2pt}
\setlength\belowdisplayskip{2pt}
	\begin{aligned}
	S_+ &= \sum\limits_{k = 1}^{\left\lfloor {\frac{{( \underline N- 1)\lambda \eta \cos \nu }}{{4l}}} \right\rfloor } {\sum\limits_{\delta  = \left\lceil {\frac{{4lk}}{{\lambda \eta \cos \nu }} - t} \right\rceil }^{\left\lfloor {\frac{{4lk}}{{\lambda \eta \cos \nu }} + t} \right\rfloor } {( \underline N- \delta )} } \\
		&= \sum\limits_{k = 1}^{\left\lfloor {\frac{{( \underline N- 1)\lambda \eta \cos \nu }}{{4l}}} \right\rfloor } {\left( {2\underline N - \left\lceil {\frac{{4lk}}{{\lambda \eta \cos \nu }} - t} \right\rceil  - \left\lfloor {\frac{{4lk}}{{\lambda \eta \cos \nu }} + t} \right\rfloor } \right)} \\
		&\times \left( {\left\lfloor {\frac{{4lk}}{{\lambda \eta \cos \nu }} + t} \right\rfloor  - \left\lceil {\frac{{4lk}}{{\lambda \eta \cos \nu }} - t} \right\rceil  + 1} \right).
	\end{aligned}
	\label{int_plus}
\end{equation}
Thus, for $\eta  \ge \frac{{4l}}{{(\underline{N}- 1)\lambda \cos \nu }},$ $S$ in (\ref{S}) reduces to 
\begin{equation}
	\begin{aligned}
		S={G_L}{\underline N^2} + ({G_H} - {G_L})(S_0+2S_+).
	\end{aligned}
\label{grating}
\end{equation}

Besides, the value of $t$ determines the number of $\Delta$ within the main lobe or grating lobes. Specifically, when $t<1$, i.e., $\eta>\frac{{4\alpha l}}{{\lambda \overline N\cos \nu }},$ the maximal number of $\Delta$ within the main lobe or grating lobes will always be $1$. In the following, we will discuss three cases based on different values of $\eta$.

\emph{Case 1: $\eta  < \frac{{4l}}{{\lambda \overline N(\underline N - 1)\cos \nu }}$.} 
Since $\eta  < \frac{{4l}}{{(\underline{N}- 1)\lambda \cos \nu }},$ no grating lobe will be generated in this case, and the whole range of $\Delta\in[-(\underline{N}-1),\underline{N}-1]$ can be included in the main lobe. By substituting $\left\lfloor t \right\rfloor  = \underline {N} - 1$ into (\ref{int_0}), $S_0$ can be further expressed as
\begin{equation}
	\begin{aligned}
		S_0 =  { - {{(\underline {N} - 1)}^2} + (2 - 1)(\underline {N} - 1) +\underline {N} } = {\underline {N}^2}.
	\end{aligned}
\end{equation}
Based on (\ref{nograting}), the EDoF $\varepsilon$ reduces to 
\begin{equation}
	\setlength\abovedisplayskip{2pt}
\setlength\belowdisplayskip{2pt}
	\begin{aligned}
		\varepsilon  &= \frac{{{{\bar N}^2}{\underline {N}^2}}}{{G_L}{\underline {N}^2} + ({G_H} - {G_L})S_0}=\frac{{{{\bar N}^2}{\underline {N}^2}}}{{{G_L}{\underline {N}^2} + ({G_H} - {G_L}){\underline {N}^2}}} \\
		&= \frac{{{{\bar N}^2}}}{{{G_H}}} \approx 1.
	\end{aligned}
\end{equation}

\emph{Case 2: $\frac{{4l}}{{\lambda \overline N(\underline N - 1)\cos \nu }} \le \eta  < \frac{{4\alpha l}}{{\lambda \overline N\cos \nu }}.$}
Since $\eta  < \frac{{4\alpha l}}{\lambda \overline N\cos \nu}<\frac{{4l}}{{(\underline{N}- 1)\lambda \cos \nu }}$. Based on (\ref{int_0}) and (\ref{nograting}), we have 
\begin{equation}
	\setlength\abovedisplayskip{2pt}
\setlength\belowdisplayskip{2pt}
	\begin{aligned}
		S=&{G_L}{\underline N^2} + ({G_H} - {G_L})S_0\\
		=&{G_L}{{\underline N}^2}+\left( {{G_H} - {G_L}} \right)\left( { - {{\left\lfloor t \right\rfloor }^2} + (2\underline N - 1)\left\lfloor t \right\rfloor  + \underline N} \right).
		\label{s2}
	\end{aligned}
\end{equation}
 Therefore, $\varepsilon$ can be written as
\begin{equation}
	\setlength\abovedisplayskip{2pt}
\setlength\belowdisplayskip{2pt}
	\begin{aligned}
		\varepsilon = \frac{{{{\overline N}^2}{{\underline N}^2}}}{\left( {{G_H} - {G_L}} \right)\left( { - {{\left\lfloor t \right\rfloor }^2} + (2\underline N - 1)\left\lfloor t \right\rfloor  + \underline N} \right) + {G_L}{{\underline N}^2}}.
	\end{aligned}
\end{equation}

\emph{Case 3: $\frac{{4\alpha l}}{{\lambda \bar N\cos \nu }} \le \eta  < \frac{{4l}}{{(\underline{N} - 1)\lambda \cos \nu }}$.}
In this case, no grating lobe will be generated and we have $t<1$, thus yielding $\left\lfloor t \right\rfloor  = 0.$ By substituting $\left\lfloor t \right\rfloor  = 0$ into (\ref{int_0}), we obtain $S_0=\underline{N}$, and 
\begin{equation}
	\begin{aligned}
		S={G_L}{\underline N^2} + ({G_H} - {G_L})\underline{N}
		=G_H\underline{N}+G_L\underline{N}(\underline{N}-1).
	\end{aligned}
\end{equation}
Thus, $\varepsilon$ can be expressed as
\begin{equation}
	\begin{aligned}
		\varepsilon = \frac{{{{\overline N}^2}{{\underline N}^2}}}{G_H\underline{N}+G_L\underline{N}(\underline{N}-1)} = \frac{{{{\overline N}^2}\underline N}}{G_H+G_L(\underline{N}-1)}.
	\end{aligned}
\label{case3}
\end{equation}

\emph{Case 4: $\eta  \ge \frac{{4l}}{{( \underline{N}- 1)\lambda \cos \nu }}$.} 
In this case, grating lobes may exist with the increase of $\eta$. However, since $\alpha\ll\overline{N}$, we always have $t=\frac{{4\alpha l}}{{\lambda \bar N\eta \cos \nu }}<1\ll {\frac{{4l}}{{\lambda \eta \cos \nu }}}$. This indicates that $S_+$ in (\ref{int_plus}) can be reduced to 
\begin{equation}
	\begin{aligned}
			{S_ + } &\approx \sum\limits_{k = 1}^{\left\lfloor {\frac{{( \underline{N}- 1)\lambda \eta \cos \nu }}{{4l}}} \right\rfloor } {\left( {2\underline{N} - \left\lceil {\frac{{4lk}}{{\lambda \eta \cos \nu }}} \right\rceil  - \left\lfloor {\frac{{4lk}}{{\lambda \eta \cos \nu }}} \right\rfloor } \right)} \\
			&\times \left( {\left\lfloor {\frac{{4lk}}{{\lambda \eta \cos \nu }}} \right\rfloor  - \left\lceil {\frac{{4lk}}{{\lambda \eta \cos \nu }}} \right\rceil  + 1} \right)\\
			&\mathop  \approx \limits^{(b)}  0,
	\end{aligned}
\label{case4}
\end{equation}
where $(b)$ holds when $\frac{{4lk}}{{\lambda \eta \cos \nu }}$ is not an integer. Note that the possibility for $\frac{{4lk}}{{\lambda \eta \cos \nu }}$ being an integer is small. (\ref{case4}) shows that the probability that no $\Delta$ will be included in each grating lobe increases as the grating lobes become narrower. Therefore, the contribution of grating lobes to $\sum\nolimits_{\Delta  \in {{\rm{S}}_{\rm{H}}}} {w(\Delta )} $ can be neglected, and the expression of $\varepsilon$ is the same as (\ref{case3}) in Case 3. 

Furthermore, by substituting $t=\frac{4\alpha l}{\lambda \overline{N}\eta \cos \nu }$ into $\varepsilon$, we can obtain the closed-form expression of EDoF given by (\ref{final-dof}).

\section*{APPENDIX C: Proof of Theorem \ref{antenna_selection}}
For single-user far-field MIMO communications, the EDoF is independent of $\eta$. 

For single-user near-field MIMO communications, the EDoF can be expressed as a piecewise function in (\ref{final-dof}). 

When $\eta  < \frac{{4l}}{{\lambda \overline N(\underline N - 1)\cos \nu }},$ we have $\varepsilon = \frac{{{\overline{N}^2}}}{{{G_H}}}$. Since $G_H\approx \overline{N}^2,$ $\varepsilon$ can be reduced to $\varepsilon \approx 1.$ In this case, $\varepsilon$ is independent of $\eta$.

When $\frac{{4l}}{{\lambda \overline N(\underline N - 1)\cos \nu }} \le \eta  < \frac{{4\alpha l}}{{\lambda \overline N\cos \nu }},$ $\varepsilon$ can be expressed as
\begin{equation}
	\small
\begin{aligned}
\varepsilon (t) =& \frac{{{{\overline N}^2}{{\underline N}^2}}}{\left( {{G_H} - {G_L}} \right)\left( { - {{\left\lfloor t \right\rfloor }^2} + (2\underline N - 1)\left\lfloor t \right\rfloor  + \underline N} \right) + {G_L}{{\underline N}^2}}\\
=&\frac{{{\overline N}^2}{{\underline N}^2}}{\Gamma(t)},
\end{aligned}
\end{equation}
where $\Gamma(t)\triangleq{ {\left( {{G_H} - {G_L}} \right)}(- {t^2} + (2\underline{N} - 1)t + \underline{N})+G_L\underline{N}^2}.$ Since $t=\frac{4\alpha l}{\lambda \overline{N}\eta \cos \nu }>1$, we have $1 \le t < \alpha (\underline N - 1)$. Since $\underline{N}\ge 2$ and $\alpha\in[0,1]$, it can be easily seen that $\Gamma(t)$ increases monotonically with $t$ when $1 \le t < \alpha (\underline N - 1)<\underline{N}-\frac{1}{2}$, which means  $\Gamma(\eta)$ decreases monotonically with the increase of $\eta$. Therefore, for this case, the EDoF $\varepsilon$ increases monotonically with $\eta$.

When $\eta  \ge \frac{{4\alpha l}}{{\lambda \overline N\cos \nu }},$ we have $\varepsilon = \frac{{{{\overline N}^2}\underline N}}{{{G_H}+G_L(\underline{N}-1)}}$. Since $G_H\approx\overline{N}^2\gg G_L$, we have $\varepsilon\approx\underline{N}$, which is a constant independent of $\eta$.

\bibliographystyle{IEEEtran}
\bibliography{IEEEabrv,document}
\end{document}